\documentclass{aa}
\usepackage{graphicx}
\usepackage{natbib}
\bibpunct{(}{)}{;}{a}{}{,}
\usepackage{txfonts}
\usepackage{longtable}

\begin{document}
\title{Chemical composition and constraints on mass loss for globular clusters in dwarf galaxies: \object{WLM} and \object{IKN}\thanks{
Based on observations made with ESO telescopes at the La Silla Paranal Observatory under programme ID 077.B-0354(A), and at the W. M. Keck Observatory, which is operated as a partnership among the California Institute of Technology, the University of California and the National Aeronautics and Space Administration.}
}

\author{S{\o}ren S. Larsen \inst{1} \and
  Jean P. Brodie \inst{2} \and
  Duncan A. Forbes \inst{3} \and
  Jay Strader \inst{4}
}
\institute{Department of Astrophysics/IMAPP, Radboud University Nijmegen, PO Box 9010, 6500 GL Nijmegen, The Netherlands \\ email: \texttt{s.larsen@astro.ru.nl}
\and
  UCO/Lick Observatory, University of California, Santa Cruz, CA 95064, USA
\and
  Centre for Astrophysics \& Supercomputing, Swinburne University of Technology,  VIC 3122 Hawthorn, Australia
\and
  Department of Physics and Astronomy, Michigan State University, East Lansing, Michigan 48824, USA
}


\date{Received 14 September 2013 / Accepted 5 April 2014}

\abstract
{}
{We determine the metallicities of globular clusters (GCs) in the \object{WLM} and \object{IKN} dwarf galaxies, using VLT/UVES and Keck/ESI spectroscopy. These measurements are combined with literature data for field stars to constrain GC formation scenarios. For the \object{WLM GC}, we also measure detailed abundance ratios for a number of light, $\alpha$, Fe-peak, and $n$-capture elements, which are compared with literature data for the Fornax dSph and the Milky Way.
}
{
The abundances are derived by computing synthetic integrated-light model spectra and adjusting the input composition until the best fits to the observed spectra are obtained. 
}
{
We find low metallicities of $\mathrm{[Fe/H]}=-2.0$ and $-2.1$ for the \object{WLM GC} and the GC \object{IKN-5}, respectively. We estimate that 17\%--31\% of the stars with $\mathrm{[Fe/H]}\le-2$ in WLM belong to the GC, and \object{IKN-5} may even contain a similar number of metal-poor stars as the whole of the IKN dwarf itself.
While these fractions are much higher than in the Milky Way halo, we have previously found a similarly high ratio of metal-poor GCs to field stars  in the \object{Fornax dSph}. The overall abundance patterns in the WLM GC are similar to those observed for GCs in the Fornax dSph: the [Ca/Fe] and [Ti/Fe]  ratios are super-Solar at about $+0.3$ dex, while [Mg/Fe] is less elevated than [Ca/Fe] and [Ti/Fe]. The [Na/Fe] ratio is similar to the averaged [Na/Fe] ratios in Milky Way GCs, but higher (by $\sim2\sigma$) than those of Milky Way halo stars. Iron-peak (Mn, Sc, Cr) and heavy elements (Ba, Y, La) generally follow the trends seen in the Milky Way halo.
}
{The GCs in the \object{WLM} and \object{IKN} dwarf galaxies resemble those in the \object{Fornax dSph} by being significantly more metal-poor than a typical halo GC in the Milky Way and other large galaxies. They are also substantially more metal-poor than the bulk of the field stars in their parent galaxies.
It appears that only a small fraction of the \object{Milky Way} GC system could have been accreted from galaxies similar to these dwarfs.  
The relatively high Na abundance in the \object{WLM GC} suggests that the [Na/O] anti-correlation is present in this cluster, while the high ratios of metal-poor GCs to field stars in the dwarfs are in tension with GC formation scenarios that require GCs to have lost a very large fraction of their initial mass.
}


\titlerunning{High-dispersion spectroscopy of GCs in WLM and IKN}
\authorrunning{S. S. Larsen et al.}
\maketitle

\section{Introduction}

\defcitealias{Larsen2012}{L12b}
\defcitealias{Larsen2012a}{L12a}

The formation of globular star clusters (GCs) was evidently a common phenomenon in the early Universe. The ``richness'' of the GC population associated with a galaxy is conveniently expressed in terms of the GC specific frequency, $S_N = N_\mathrm{GC} \times 10^{0.4 (15 + M_V)}$, where $M_V$ is the host galaxy absolute magnitude and $N_\mathrm{GC}$ is the number of GCs \citep{Harris1981}.
Most large galaxies have $S_N$ of the order of unity to a few, 
but there are significant variations both within galaxies and from one galaxy to another. Two notable trends are that (1) the GC/field star number ratio tends to increase with decreasing metallicity within a given galaxy \citep{Forte1981,Forbes1997,Larsen2001,Forbes2001,Harris2002,Harris2007} and that (2) many dwarf galaxies host GCs in disproportionately large numbers compared to larger galaxies \citep{Miller2007,Peng2008,Georgiev2010,Harris2013}. 

The total integrated luminosity of the Milky Way GC system is about $1.5\times10^7 L_{V\odot}$ \cite[using data from][]{Harris1996}, which corresponds to a mass of $2.2\times10^7 M_\odot$ for an average visual mass-to-light ratio of $\Upsilon_V = 1.45$ \citep{McLaughlin2000}. Assuming that about two thirds of these GCs are associated with the halo \citep[e.g.][]{Zinn1985}, the GC system then accounts for 1\%--2\% of the stellar halo mass \citep{Suntzeff1991}.
This \emph{present-day} ratio almost certainly differs significantly from the corresponding ratio at the time of the formation of the GCs or shortly thereafter. All star clusters evolve dynamically, and for GCs with initial masses of the order of $10^5 M_\odot$, the time scale for dissolution due to the combined effects of two-body relaxation and bulge/disc shocks is comparable to the Hubble time  \citep{Fall1977,Fall2001,Jordan2007,Kruijssen2009}.  Clusters that formed with initial masses below this limit in the early Universe will thus have dissolved by now and their stars become part of the general halo field star population.
The mass lost from the GC system due to disruption may be comparable to the present-day stellar mass of the Galactic halo \citep{PortegiesZwart2010}.

The amount and rate of mass loss experienced earlier in the lifetime of GCs, when they may have interacted more strongly with the surrounding interstellar medium, remains more uncertain. 
Galactic star-forming regions exhibit a great deal of hierarchical structure and ``clustering'', but the fraction of stars that end up as members of bound, long-lived clusters is only of the order of a few percent in normal galaxies, though possibly higher in starburst environments  \citep{Fall2004,Goddard2010,Lada1991,Lada2003,Larsen2000a,Silva-Villa2011,Kruijssen2012}. This has led to the notion of 
 ``infant mortality'' or ``infant weight loss'', whereby young star clusters (or parts thereof) may become unbound after the expulsion of left-over gas from star formation \citep{Hills1980,Elmegreen1983,Goodwin1997,Boily2003,Goodwin2006,Fall2010}. 
 
Large amounts of early mass loss have also been suggested as a solution to the ``mass budget'' problem, which is inherent to many theories for the origin of chemical abundance anomalies observed in GCs. Compared to field stars, GC stars display anomalous abundance patterns of many light elements \citep{Carretta2009}, and it has been suggested that this may be due to self-pollution within the clusters via mass loss from massive asymptotic giant branch (AGB) stars or fast-rotating massive main sequence stars (``spin stars''). However, the amount of processed material returned by such stars is insufficient to account for the very large observed fractions of stars with anomalous abundance patterns, which has prompted the suggestion that a large fraction of the normal ``first-generation'' stars have been lost from the clusters. In these scenarios, the present-day GCs would be the surviving remnants of systems that were initially far more massive by perhaps a factor of 10 or more \citep{DErcole2008,Schaerer2011,Bekki2011,Valcarce2011}. Recently, scenarios have been proposed in which the anomalies arise from self-pollution during the formation of the cluster -- for example, by material released from massive interacting binaries \citep{Bastian2013a,Denissenkov2013}. The mass budget problem is less severe in these scenarios.

An interesting constraint on these ideas comes from observations of GCs in dwarf galaxies. In particular, the chemical composition has now been studied in detail for all five GCs in the \object{Fornax dSph} using high-dispersion spectroscopy of individual stars \citep{Letarte2006} and integrated light \citep[hereafter L12a and L12b]{Larsen2012a,Larsen2012}. Four of the five clusters in the Fornax dwarf are very metal-poor with metallicities of $\mbox{[Fe/H]}<-2$ (The remaining cluster, \object{Fornax 4}, has $\mbox{[Fe/H]}\approx-1.4$.).
Combined with the metallicity distribution of the field stars \citep{Battaglia2006}, this implies that a very large fraction of all the metal-poor stars in the Fornax dSph (20\%--25\%) belong to the four metal-poor GCs \citepalias{Larsen2012}. 
There is an evident tension between the relatively small number of metal-poor field stars in Fornax and scenarios that require GCs to have lost 90\% or more of their initial masses to explain multiple generations of stars \citep[but see][]{DAntona2013}.

Another consequence of the very low metallicities of the Fornax GCs is that only a small fraction of the Milky Way GC system could have been accreted from dwarf galaxies similar to the Fornax dSph. In the Milky Way, only 11 GCs or about 7\% of the GC population have $\mathrm{[Fe/H]}<-2$ \citep{Harris1996}. 
This, of course, does not necessarily exclude that the halo was built up from smaller fragments, but it does suggest that these fragments experienced a higher degree of early chemical enrichment before GCs were formed, as compared to the surviving dwarf galaxies that are observed today.

While the results for the Fornax dSph are interesting, it would clearly be desirable to extend this analysis to additional galaxies. In this paper, we analyse integrated-light spectroscopy of the single globular cluster known in the \object{Wolf-Lundmark-Melotte} (\object{WLM})  galaxy and the brightest of the five GCs found in the \object{IKN} dwarf spheroidal galaxy in the M81 group \citep{Georgiev2010}. Apart from determining the overall metallicities of the GCs and comparing them with the field star metallicity distributions, we also carry out a detailed abundance analysis of the \object{WLM GC} and compare our results with the data for GCs in the \object{Fornax dSph} and the Milky Way. We use the same integrated-light spectral analysis technique that was developed and applied to the Fornax GCs in \citetalias{Larsen2012a} with a few modifications described below (Sect.~\ref{sec:analysis}).

\section{Data}
\label{sec:data}

\subsection{The WLM galaxy}

\begin{figure}
\includegraphics[width=\columnwidth]{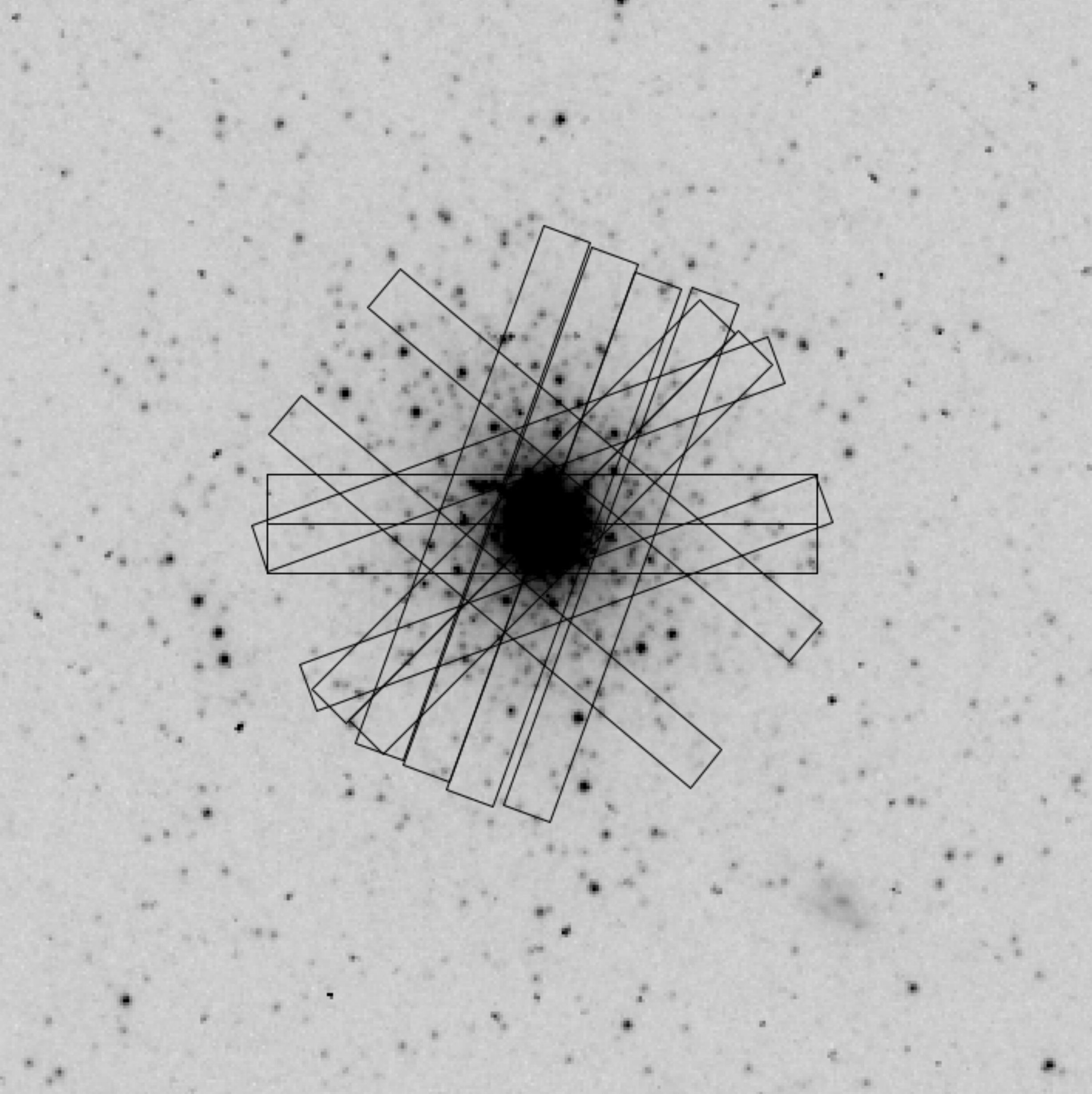}
\caption{\label{fig:uves-wlm}The UVES slit locations superimposed on an HST/WFPC2 image of the WLM globular cluster. The length of the UVES slit is $8\farcs9$. North is up and east to the left.}
\end{figure}

\begin{figure}
\includegraphics[width=\columnwidth]{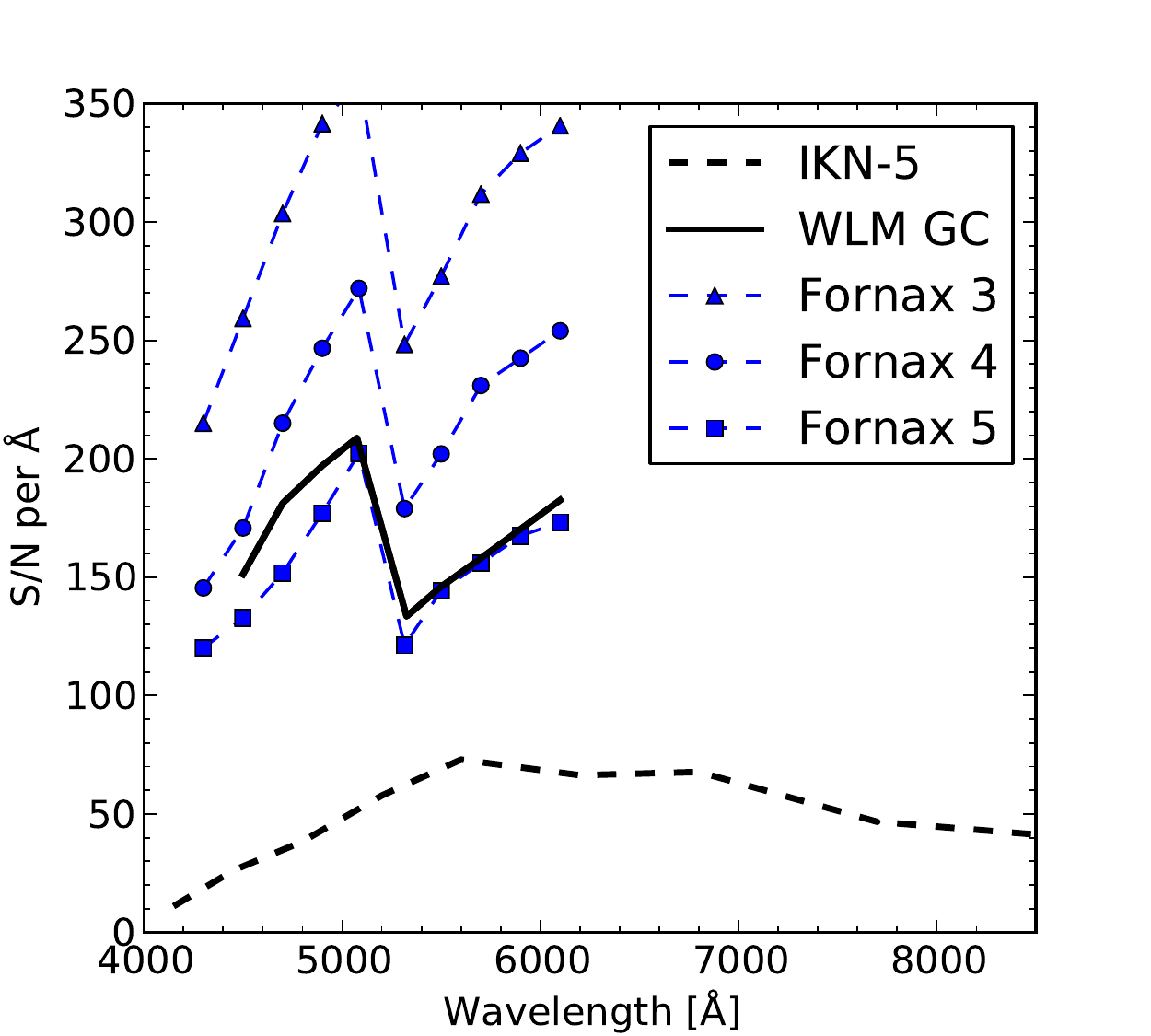}
\caption{\label{fig:s2n}Signal-to-noise ratio of the spectra.  Lines connect averages over 200 \AA\ bins. The discontinuity at 5200 \AA\ in the UVES spectra is caused by the different detectors. For reference, we have also included the GCs in the Fornax dSph \citepalias{Larsen2012a}.}
\end{figure}

The \object{WLM} galaxy is a dwarf irregular galaxy at the outskirts of the Local Group. It is located at a distance of 966 kpc \citep{Gieren2008} and hosts a single old globular cluster.
Given that late-type galaxies typically have GC specific frequencies of  $S_N\approx1$ \citep{Harris1991}, it is not unexpected to find a GC in WLM, which has an absolute magnitude of  $M_V = -14.5$ \citep{Mateo1998}. However, the \object{WLM GC} is significantly brighter than an average GC. The apparent visual magnitude, $V=16.06$ \citep{Sandage1985}, corresponds to $M_V=-9.0$ for $A_V=0.104$ \citep[][via the NASA/IPAC Extragalactic Database]{Schlafly2011}, as compared to the average $M_V\approx-7.5$ for GCs in a wide variety of galaxies \citep[e.g.,][]{Rejkuba2012}. 
 A colour-magnitude diagram of the cluster was obtained from Hubble Space Telescope (HST) photometry by \citet{Hodge1999}, who concluded that it is very old and has a low metallicity ($\mathrm{[Fe/H]}=-1.52\pm0.08$). 
From integrated-light spectroscopy, \citet{Colucci2011} found  $\mathrm{[Fe/H]}=-1.71\pm0.03$. 
The cluster has a half-light radius of about $0\farcs75$ or 3.5 pc and is significantly elongated, although no evidence for rotation has been found \citep{Stephens2006}.

We obtained integrated-light spectra of the WLM GC with the UVES spectrograph \citep{Dekker2000} on the ESO Very Large Telescope.  The observations were made in service mode on 30 July, 2 August, and 3 August 2006.
The UVES setup was similar to that used for the Fornax GCs \citepalias{Larsen2012a}, except that a slightly narrower slit ($0\farcs8$ instead of $1\arcsec$) was used, yielding a formal resolving power of $R=50\, 000$ over the wavelength range 4200 \AA\ -- 6200 \AA\ with a gap at 5150 \AA\ -- 5250 \AA\ between the two CCD detectors. For the blue detector, the pixel scale of the reduced spectra is 0.026 \AA\ pixel$^{-1}$, while it is 0.032 \AA\ pixel$^{-1}$ for the red detector.
Due to the greater distance of WLM, we did not scan the UVES slit across the cluster as was done for the Fornax GCs, but instead obtained a number of fixed pointings with the slit at different position angles and locations with respect to the cluster centre (Fig.~\ref{fig:uves-wlm}). We obtained 12 exposures, each with an integration time of 1475 s, for a total of 4 hr 55 min.

The spectra were reduced with version 5.1.0 of the UVES pipeline provided by ESO. The pipeline performs bias subtraction, flatfielding, and wavelength calibration of the spectra and merges the echelle orders to a single spectrum per detector. The 1D spectra were extracted by the pipeline using optimal weighting of the pixels across the spatial profile.
The individual 1D spectra were then co-added to produce a final single 1D spectrum used in the further analysis.
As a check we also extracted the 2D spectra and co-added the pixels in the spatial direction manually within IRAF. The $8\farcs9$ long slit was mapped onto 22 pixels in the spatial direction. We estimated the sky background by median filtering the outermost rows of pixels in the 2D spectrum with a $101\times1$ pixels median filter. The background was then subtracted from the remaining 20 pixels in each wavelength bin by linear interpolation between the outermost pixels. Finally, the 20 rows were  co-added to produce a single 1D spectrum per exposure. Analysis of these manually extracted spectra yielded similar results to the optimally extracted spectra but at slightly lower S/N. The optimally extracted spectra were used for further analysis in this paper.

\subsection{The IKN galaxy}

The \object{IKN} galaxy is a dwarf spheroidal galaxy in the \object{M81} group. It has a distance of 3.7 Mpc and a projected separation of 84 kpc from M81 \citep{Karachentsev2002}. The integrated absolute $V$ magnitude is estimated to be $M_V\approx-11.5$ \citep{Georgiev2009}, but accurate  photometry of IKN is difficult due to the presence of a very bright, nearby foreground star. From HST imaging, \citet{Georgiev2009} identified five GCs in IKN with magnitudes between $M_V=-6.7$ and $M_V=-8.5$. The GC system of IKN thus appears similar to that of the Fornax dSph, even though IKN itself is nearly two magnitudes fainter. This leads to an extremely high GC specific frequency of $S_N=125$. 
The metallicity distribution of the field stars in IKN has been determined by \citet{Lianou2010} from HST photometry of red giant branch (RGB) stars. It appears similar to that of the Fornax dSph with a broad peak around $\mathrm{[Fe/H]}\approx-1.5$ to $-1$ and with a tail extending to lower metallicities. It is thus of interest to compare the metallicities of the field stars with those of the GCs.

We observed the brightest of the five GCs \citep[IKN-5 in the list of][]{Georgiev2009} with the Echellette Spectrograph and Imager \citep[ESI;][]{Sheinis2002} on the Keck II telescope on 5 March 2013. Weather conditions were clear with a seeing around $1\arcsec$.
We obtained seven exposures of 20 min each with ESI in the echellette (cross-dispersed) mode using a $0\farcs5$ slit. The observations covered the wavelength range 3900 \AA\ -- 1.1 $\mu$m at a resolving power of $R\approx8000$, although we did not make use of the data below $\sim4600$ \AA\ where the S/N ratio becomes too low (Fig.~\ref{fig:s2n}).
The pixel scale in the dispersion direction  is $\sim0.2$ \AA\ pixel$^{-1}$.
At the distance of IKN, most globular clusters are point-like in seeing-limited ground-based observations, and the spectra were reduced with the standard MAKEE package written by T.\ Barlow.
We also attempted to observe two of the fainter clusters in IKN (IKN-2 and IKN-4), but the S/N of these data turned out to be insufficient for further analysis.
 
\section{Analysis}
\label{sec:analysis}

\begin{figure*}
\includegraphics[width=\textwidth]{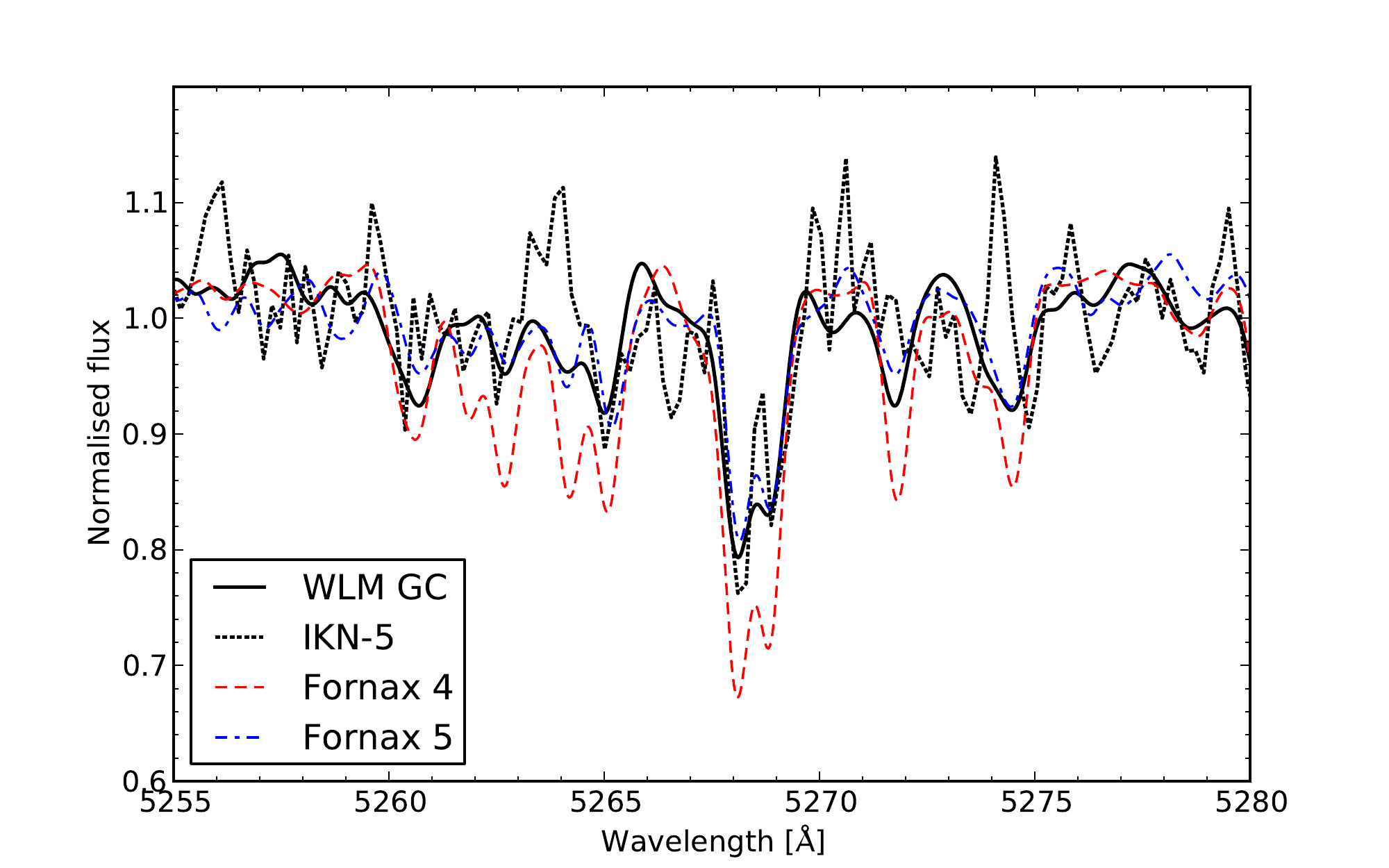}
\caption{\label{fig:fe5270}Comparison of the spectra of IKN-5 and the WLM GC with the spectra of Fornax 4 and 5 for the region near the Lick/IDS Fe5270 feature. The spectra have been smoothed with a Gaussian with $\sigma=8$ pixels, except  for the IKN-5 spectrum.}
\end{figure*}

In Fig.~\ref{fig:fe5270}, we show a small region of the \object{WLM GC} and \object{IKN-5} spectra near the Fe5270 feature. We also include the spectra of \object{Fornax 4} and \object{Fornax 5} from \citetalias{Larsen2012a}. The UVES spectra (WLM GC, Fornax 4, and  Fornax 5) have been smoothed with a Gaussian of $\sigma=8$ pixels for illustration purposes to roughly match the resolution of the \object{IKN-5} data. \citet{Larsen2012a} found metallicities of 
$\mathrm{[Fe/H]}=-1.4$ and $-2.1$ for Fornax 4 and 5, respectively.
From the comparison in Fig.~\ref{fig:fe5270}, it is already clear that the metallicities of both  IKN-5 and the WLM GC  are not very different from that of Fornax 5 and substantially lower than that of Fornax 4. 

For our analysis of the integrated-light spectra, we followed a procedure similar to that described in detail in \citetalias{Larsen2012a}. Briefly, we started by generating a Hertzsprung-Russell diagram (HRD) for each cluster. The HRD was divided into about 100 bins covering the main phases of stellar evolution (main sequence, red giant branch, and horizontal branch). For each cmd-bin, we computed a model atmosphere and a synthetic spectrum for the corresponding stellar parameters ($\log g$, $T_\mathrm{eff}$, and chemical composition) and scaled the synthetic spectrum by the number of stars in the cmd-bin. The synthetic spectra for all cmd-bins were then co-added, smoothed to the resolution of the observations, and compared with the observed integrated-light cluster spectra. The elemental abundances were adjusted and the procedure repeated until the best fit was obtained, where the ``best fit'' was defined as the fit that yielded the smallest $\chi^2$ value over a specific wavelength range.
As discussed in \citetalias{Larsen2012a}, we did not make a formal distinction between ``feature'' and ``continuum'' regions of the spectrum. The overall scaling of the model spectrum was determined by fitting the ratio of the model and observed spectra with a spline or polynomial and then multiplying the model spectrum with the fit before computing the residuals. 
We experimented with a more restrictive selection of spectral regions used for the scaling (Sect.~\ref{sec:scal}) but found no clear advantage over using the whole spectrum.
The code allows us to fit or specify fixed abundances for arbitrary combinations of elements simultaneously, although we generally fit only one at a time, starting with those having the greatest numbers of features (Fe, followed by Ti). For further details, we refer to \citetalias{Larsen2012a}.

We used the Linux versions of the \texttt{ATLAS9} and \texttt{SYNTHE} model atmosphere and spectral synthesis codes originally written by R.\ Kurucz \citep{Kurucz1970,Kurucz1979,Kurucz1981,Sbordone2004} with line lists and other data from the website of F.\ Castelli\footnote{\texttt{http://wwwuser.oat.ts.astro.it/castelli/}}. 
The Castelli line list is a modified version of the list available from the Kurucz website\footnote{\texttt{http://kurucz.harvard.edu}} \citep{Castelli2004}.
We generally started by solving for the best-fitting overall scaling of the abundances \citep[relative to the Solar composition;][]{Grevesse1998}, adopting a fixed enhancement of $+0.3$ dex for the $\alpha$-elements (O, Ne, Mg, Si, S, Ar, Ca, and Ti) relative to the Solar composition.  In this initial run, we also solved for the best-fitting broadening of the model spectra. In subsequent iterations, we then solved for the abundances of  individual elements, using custom-defined wavelength ranges that contained features from the corresponding elements. For the IKN-5 spectrum, only limited information about individual abundances could be extracted due to the lower S/N and spectral resolution, and in this paper, we only discuss the Fe abundance of this cluster.

In our previous analysis of the Fornax GCs, the input HRDs were based mainly on the empirical colour-magnitude diagrams (CMDs) available from HST data. The HRDs were thus known \emph{a priori}. While the WLM GC does have a published CMD, which is useful for constraining the horizontal branch morphology, it is not nearly as deep as those available for the Fornax GCs (mainly due to the greater distance), and we therefore had to rely more heavily on theoretical models.
To this end, we used model isochrones and luminosity functions from \citet{Dotter2007}. The stellar masses were assumed to be distributed according to a power-law, $\mathrm{d}N/\mathrm{d}M \propto M^{-\alpha}$, with the \citet{Salpeter1955}  slope, $\alpha = 2.35$, down to a lower mass limit of $0.4 M_\odot$. The actual shape of the low-mass end of the stellar mass function in GCs may be strongly influenced by dynamical evolution, which leads to the preferential loss of low-mass stars \citep{Kruijssen2009,DeMarchi2010}. These stars only contribute a few percent of the integrated light, and the effects on the spectra are thus relatively small. The resulting uncertainties are probably less than $\sim0.1$ dex for most individual elements \citepalias{Larsen2012a}.
We combined the Dotter et al.\ models for the main sequence (MS), sub-giant branch (SGB), and RGB with empirical data for the horizontal branch (HB), using observations from the ACS survey of Galactic GCs \citep{Sarajedini2007}.  
The weighting of the empirical HB data was determined based on the number of RGB stars in the range $1 < M_V < 2$ in the model HRDs and in the empirical ACS data. The stellar parameters for the MS, SGB, and RGB stars were taken directly from the models, while they were derived for the HB stars from the ACS photometry as described in \citetalias{Larsen2012a}. 

A complication of this modified approach is that we did not a priori know the metallicities of the clusters, which in turn were needed to select the proper isochrones for the integrated-light analysis. We therefore started by constructing  model HRDs based on rough initial guesses for the metallicities. We could then use these initial model HRDs to obtain  improved estimates of the cluster metallicities based on the integrated-light spectra, adjust our choices of model isochrones, and so on. The metallicities derived from the integrated spectra tend to be \emph{overestimated} if the metallicities of the input isochrones are \emph{too low} and vice versa (Sect.~\ref{sec:uncertainties}), and this procedure generally converged towards stable estimates of the  metallicities after a couple of iterations.

\begin{figure}
\centering
\includegraphics[width=\columnwidth]{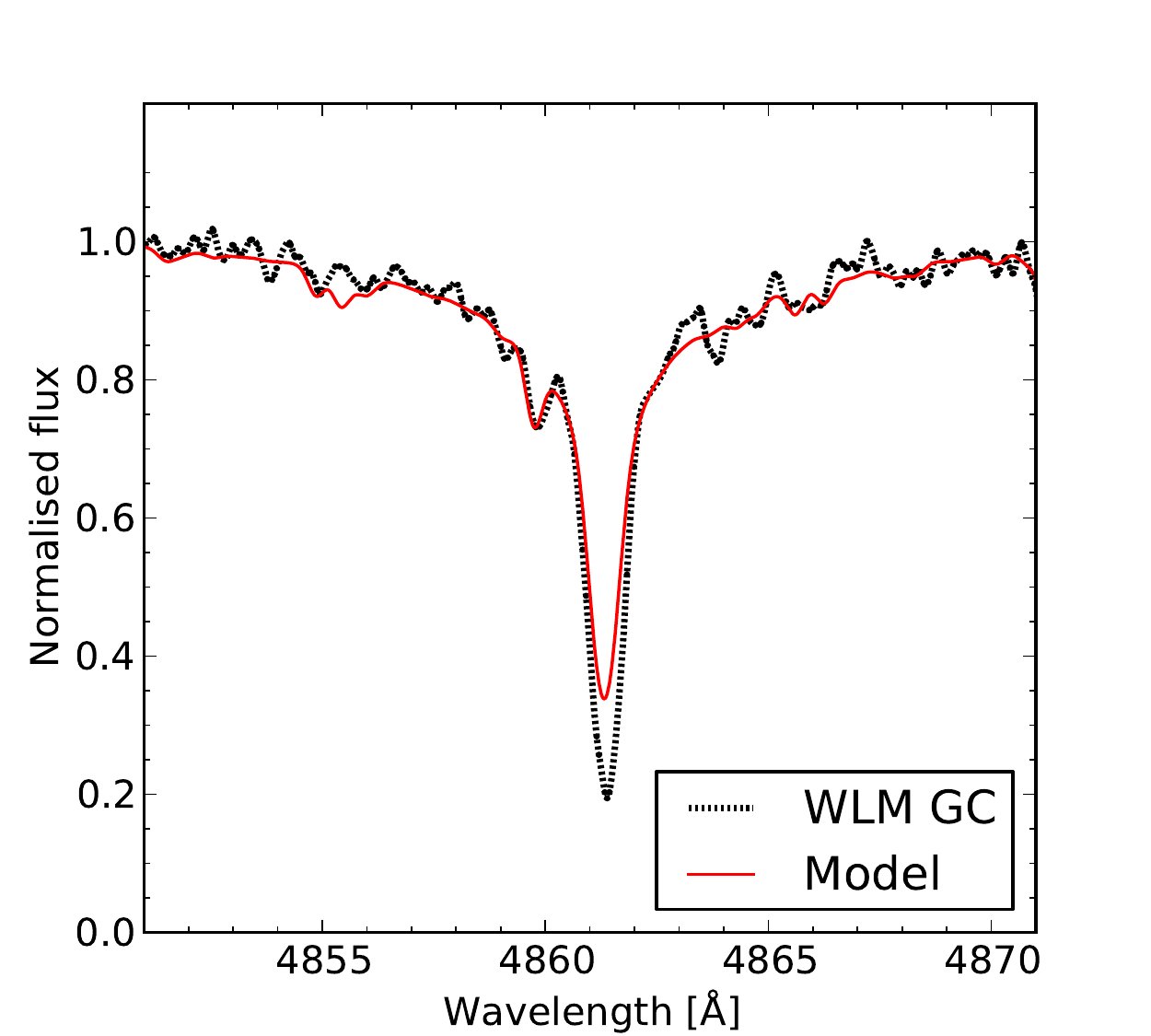}
\caption{\label{fig:hbcmp}Comparison of the WLM GC and model spectrum around the H$\beta$ line. The model and observed spectra have both been smoothed with a Gaussian kernel with $\sigma=3$ pixels.}
\end{figure}

A more difficult problem was to determine the appropriate ages of the model isochrones. 
In principle these may also be obtained as a free parameter in the model fits, since different spectral features have different relative sensitivities to age and chemical composition. For example, the hydrogen Balmer lines are commonly used as age indicators in the Lick/IDS system. However, this approach is not without complications, as other parameters than age can affect the Balmer lines \citep[e.g.][]{Poole2010}. Moreover, the cores of the Balmer lines are affected by non-LTE effects that make the observed cores deeper than those produced by LTE models \citep{Auer1970,Barklem2007}. This is clearly seen in our data: as an example, Fig.~\ref{fig:hbcmp} compares the best-fitting model spectrum for the WLM GC with the observed spectrum. While the wings of the model H$\beta$ line match the data well,  the core is clearly too shallow. Changing the age of the input isochrone within reasonable limits (several Gyr) does not significantly affect this comparison; the core does not get significantly deeper even for much younger ages. 
In principle, other lines may also be used. \citet{Colucci2011a} determined ages for clusters in the Large Magellanic Cloud (LMC) by searching for the age where different Fe lines gave the most consistent set of abundances. In this way,  the LMC globular clusters in their sample were found to have ages older than 5--7 Gyr, which is consistent with other determinations but with large uncertainties.
Here, we have opted \emph{not} to solve for the age but instead limited ourselves to quantifying the sensitivity of our results to age variations by repeating the analysis for a few different ages (see Sect.~\ref{sec:uncertainties}). 

In the following, we comment on various aspects of the analysis specific to the individual clusters.

\subsection{\object{WLM GC}}
\label{sec:wlmgc}

The similarity of the \object{Fornax 5} and \object{WLM GC} spectra offered an opportunity to compare the analysis based on an empirical CMD with  the more theoretical approach adopted here. 
We started by using the Fornax 5 CMD from \citetalias{Larsen2012a} as input for the integrated-light analysis of the WLM GC. From this, we obtained an iron abundance of $\mathrm{[Fe/H]}=-1.97$. 
We then generated a theoretical HRD by combining a 13 Gyr isochrone with $\mathrm{[Fe/H]}=-2.0$ with the horizontal branch of the Galactic GC \object{NGC 6779} (M56). The metallicity of \object{NGC~6779} \citep[$\mathrm{[Fe/H]}=-1.94$, according to][]{Harris1996} closely matches that of the \object{WLM GC} derived above, and \object{NGC~6779} has an extended blue HB \citep{Hatzidimitriou2004} that appears similar to that of  the  WLM GC \citep{Hodge1999}.
Using this HRD as input for the model spectra, we obtained almost exactly the same iron abundance as that derived based on the Fornax 5 CMD, $\mathrm{[Fe/H]}=-1.96$ (Sect.~\ref{sec:results}). The outcome of this  consistency check  gives us some confidence that the analysis based on theoretical HRDs gives reasonable results. Further tests of the isochrone approach are discussed below (Sect.~\ref{sec:uncertainties}).

The best-fitting Gaussian dispersion for the smoothing of the model spectra was $\sigma_\mathrm{smooth}=9.9$ km s$^{-1}$, which is determined as an average over the 200 \AA\ intervals used for the initial fits. 
Assuming that the resolving power approximately represents a Gaussian FWHM value
(as was indeed found to be the case by \citetalias{Larsen2012a}), the instrumental resolution corresponds to $\sigma_\mathrm{instr} = 2.55$ km s$^{-1}$. After subtracting this in quadrature, the line-of-sight velocity dispersion of the cluster stars is $\sigma_\mathrm{1D}=9.6$ km s$^{-1}$.

\subsection{\object{IKN-5}}

As noted above, the ESI spectrum has a lower resolution of $R\approx8000$. This is significantly lower than the spectral resolution traditionally considered necessary for detailed abundance analysis, although one can to some extent trade spectral resolution for S/N. For example, \citet{Conroy2014} have recently measured abundances of a number of individual elements in high S/N integrated-light galaxy spectra from the Sloan Digital Sky Survey (SDSS).
In these cases, it is typically not possible to identify the true continuum (particularly at higher metallicities), but this difficulty is alleviated at least partly by the full spectral fitting procedure.

Nevertheless, due to the relatively low S/N of the IKN-5 spectrum, the amount of detailed information we could extract was more limited than for the WLM GC.
Based on fits to a number of individual, stronger features, such as the Ca {\sc ii} IR and Mg {\sc i}b triplets, we found a best-fitting smoothing of about 14 km s$^{-1}$, which is slightly less than the formal resolution of the instrument ($\sigma_\mathrm{instr}\approx16$ km s$^{-1}$) if the $R$ value is assumed to correspond to a Gaussian FWHM.
Although most  lines are too weak to be measurable individually, the large number of Fe lines present throughout the optical part of the spectrum meant that we were still  able to constrain the Fe abundance of IKN-5 reasonably well.
Our best fit yielded a metallicity of $\mathrm{[Fe/H]}=-2.1$ with a weighted rms scatter of rms$_w = 0.19$ dex (see Sect.~\ref{sec:results}).
Uncertainty about the appropriate amount of smoothing has some effect on the results: the instrumental resolution ($\sigma_\mathrm{instr} \approx 16$ km s$^{-1}$) convolved with a plausible velocity dispersion of $\approx 8$ km s$^{-1}$ would yield a resolution of $\sigma_\mathrm{smooth} = 18$ km /s$^{-1}$; in this case, the Fe abundance increased to $\mathrm{[Fe/H]} = -1.95$.

We carried out several tests of our fitting algorithm to verify its performance in this less ideal case, as compared to the relatively high resolution and high S/N spectra on which it has previously been applied.
First, we reanalysed the IKN-5 spectrum after shifting the spectrum by 6 \AA, a value that is large enough so that features in the model- and observed spectra no longer match. We were thus essentially fitting a pure noise spectrum and expected the fit to converge towards a very low metallicity. Indeed, the algorithm returned [Fe/H] values at the lower boundary of the range probed for most spectral bins ($\mathrm{[Fe/H]}=-4$), which is far below the value measured in the correctly shifted IKN-5 spectrum.

As a second test, we broadened the UVES spectrum of the \object{WLM GC} with a Gaussian dispersion of 16 km s$^{-1}$, rebinned it to a resolution of 0.2 \AA\ pixel$^{-1}$, and added random Gaussian noise to each pixel corresponding to a S/N=25. This roughly mimics the characteristics of the IKN-5 spectrum. On this degraded spectrum we measured an iron abundance of $\mathrm{[Fe/H]}=-1.98$ with a weighted rms$_w$ = 0.13 dex, which closely agrees with the $\mathrm{[Fe/H]}=-1.96$ measured on the original WLM GC spectrum. From this, we conclude that no significant systematic errors on the $\mathrm{[Fe/H]}$ abundance are introduced by the lower spectral resolution and S/N of the IKN-5 spectrum.
From the same experiment, we also found that reliable measurements of abundance ratios, such as [Mg/Fe], [Ca/Fe], and [Ti/Fe] required a significantly higher S/N. We found that a S/N of about 100 was required to obtain abundances of these elements from the degraded spectrum that agreed with those measured in the original UVES spectrum within $\sim0.05$ dex. 

Finally, we tested our procedure on synthetic spectra. Using the \object{Fornax 5} CMD, we generated synthetic spectra for an input metallicity of $\mathrm{[Fe/H]}=-2$ and $[\alpha/\mathrm{Fe}]=+0.2$. These spectra were then degraded and rebinned to the  resolution of the ESI data and random noise was added, corresponding to S/N=10, 20, 30, 50, and 100 per pixel. We then applied our fitting algorithm to the synthetic spectra in 200 \AA\ bins between 4600 \AA\ and 5400 \AA, again using the Fornax 5 CMD as input. This procedure was repeated ten times for different random realisations of the synthetic spectra.
We found that even for a S/N as low as 10 per pixel (i.e., about 20 per \AA) the input metallicities were recovered with excellent accuracy: for the four wavelength bins, the fit returned average metallicities of $[\mathrm{Fe/H}]=-2.09$ ($\sigma=0.29$), $-1.96$ ($\sigma=0.21$), $-2.01$ ($\sigma=0.15$) and $-1.94$ ($\sigma=0.18$), where the numbers in the parentheses are the standard deviations of the ten realisations. The weighted average is 
$\langle[\mathrm{Fe/H}]\rangle=-1.99\pm0.03$, which is very close to the input value of $\mathrm{[Fe/H]}=-2$.

From these tests, we conclude that the ESI spectrum of \object{IKN-5} is adequate for reliably measuring the iron abundance.   
Possible systematic errors due to the lower spectral resolution are comparable to those arising from a variety of other sources \citepalias{Larsen2012a}. Individual abundance ratios are more uncertain and are not discussed further in this paper but should be measurable given sufficient S/N.

\section{Results}
\label{sec:results}

\begin{figure*}
\centering
\includegraphics[width=\textwidth]{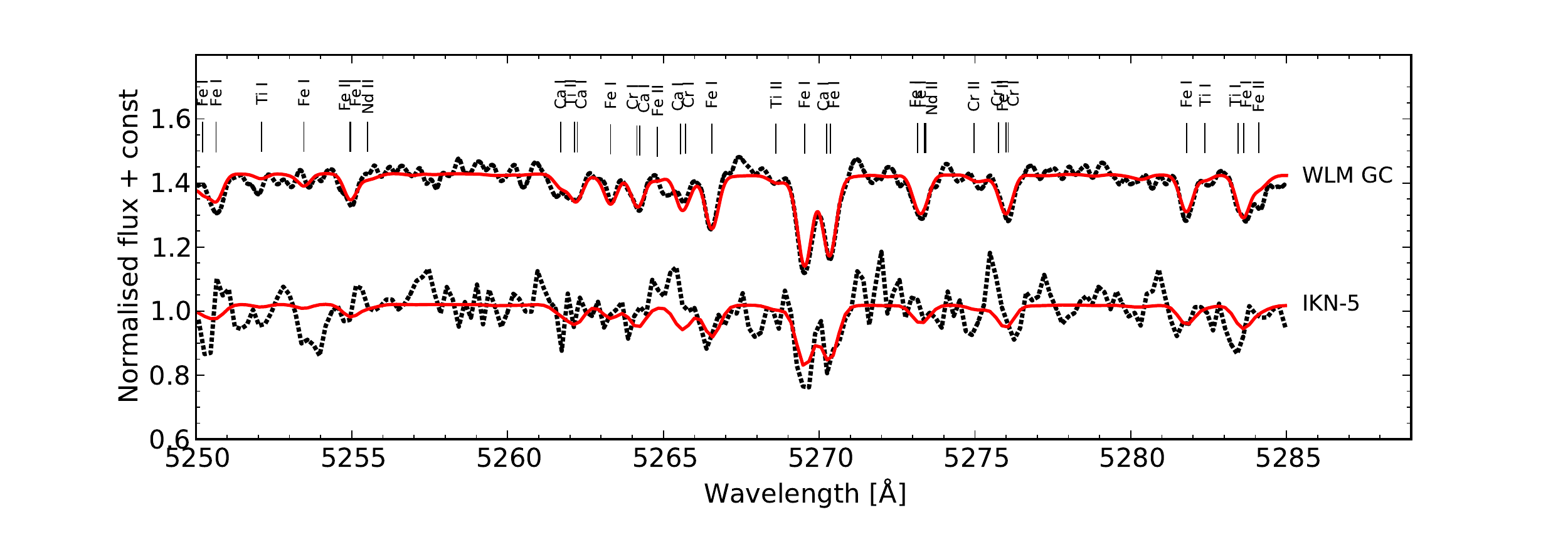}
\includegraphics[width=\textwidth]{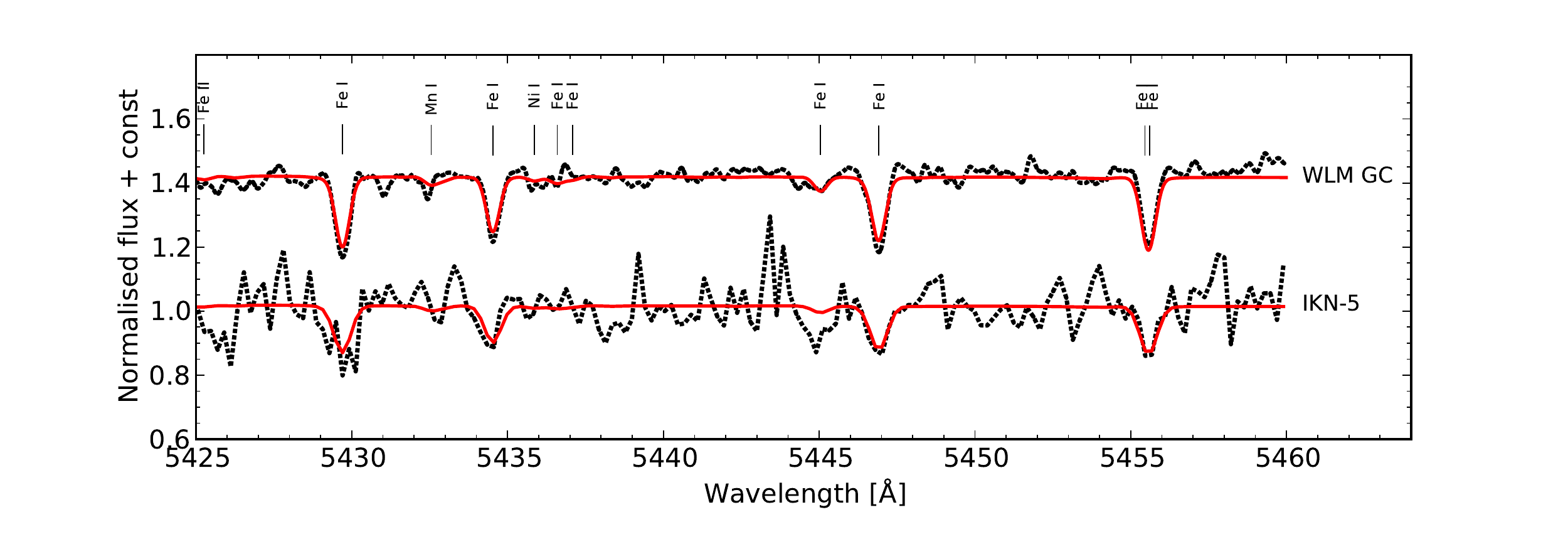}
\caption{\label{fig:ca56}Example spectral fits. The black dotted curves are the observed spectra, while the solid curves (red in the on-line version) are the best-fitting model spectra. The spectra of the WLM GC have been smoothed by a Gaussian kernel with $\sigma = 3$ pixels ($\approx0.1$\AA), while no smoothing has been applied to IKN-5.}
\end{figure*}

\onltab{1}{
\begin{longtable}{lcc}
\caption{\label{tab:wlmabun}Individual abundance measurements for the \object{WLM GC}.}\\ \hline \hline
Wavelength [\AA] & Value & Error \\ \hline
\endfirsthead
\caption{continued.}\\ \hline \hline
Wavelength [\AA] & Value & Error \\ \hline
\endhead
\hline
\endfoot
\mbox{[Fe/H]} \\
4400.0--4600.0 & $-1.923$ & 0.016 \\
4600.0--4800.0 & $-2.004$ & 0.021 \\
4800.0--5000.0 & $-2.053$ & 0.016 \\
5000.0--5150.0 & $-1.893$ & 0.015 \\
5250.0--5400.0 & $-1.853$ & 0.025 \\
5400.0--5600.0 & $-2.073$ & 0.025 \\
5600.0--5800.0 & $-2.073$ & 0.041 \\
5800.0--6000.0 & $-2.153$ & 0.070 \\
6000.0--6200.0 & $-1.833$ & 0.041 \\
\mbox{[Na/Fe]} \\
5677.0--5695.0 & $+0.231$ & 0.146 \\
\mbox{[Mg/Fe]} \\
4347.0--4357.0 & $-0.169$ & 0.256 \\
4565.0--4576.0 & $+0.201$ & 0.120 \\
4700.0--4707.0 & $+0.082$ & 0.080 \\
5523.0--5531.5 & $-0.208$ & 0.126 \\
5705.0--5715.0 & $+0.150$ & 0.210 \\
\mbox{[Ca/Fe]} \\
4420.0--4460.0 & $+0.182$ & 0.051 \\
4575.0--4591.0 & $+0.001$ & 0.090 \\
4873.0--4883.0 & $+0.501$ & 0.126 \\
5259.0--5268.0 & $+0.262$ & 0.120 \\
5580.0--5610.0 & $+0.392$ & 0.055 \\
6100.0--6175.0 & $+0.231$ & 0.036 \\
\mbox{[Sc/Fe]} \\
4290.0--4330.0 & $+0.512$ & 0.075 \\
4350.0--4440.0 & $+0.292$ & 0.070 \\
4665.0--4675.0 & $-0.039$ & 0.146 \\
5026.0--5036.0 & $+0.081$ & 0.150 \\
5521.0--5531.0 & $-0.338$ & 0.226 \\
5638.0--5690.0 & $+0.160$ & 0.081 \\
\mbox{[Ti/Fe]} \\
4292.0--4320.0 & $+0.131$ & 0.056 \\
4386.0--4420.0 & $+0.321$ & 0.051 \\
4440.0--4474.0 & $+0.281$ & 0.051 \\
4532.0--4574.0 & $+0.241$ & 0.036 \\
4587.0--4593.0 & $+0.672$ & 0.146 \\
4650.0--4715.0 & $+0.241$ & 0.051 \\
4750.0--4850.0 & $+0.341$ & 0.051 \\
4980.0--5045.0 & $+0.182$ & 0.030 \\
5152.5--5160.0 & $+0.140$ & 0.115 \\
\mbox{[Cr/Fe]} \\
4250.0--4292.0 & $-0.348$ & 0.075 \\
4520.0--4660.0 & $-0.089$ & 0.045 \\
5235.0--5330.0 & $+0.061$ & 0.061 \\
5342.0--5351.0 & $-0.278$ & 0.170 \\
5407.0--5413.0 & $+0.170$ & 0.155 \\
\mbox{[Mn/Fe]} \\
4750.0--4790.0 & $-0.458$ & 0.075 \\
6010.0--6030.0 & $-0.138$ & 0.191 \\
\mbox{[Y/Fe]} \\
4355.0--4425.0 & $+0.241$ & 0.111 \\
4879.0--4889.0 & $-0.298$ & 0.171 \\
5084.0--5094.0 & $-0.549$ & 0.250 \\
\mbox{[Ba/Fe]} \\
4551.0--4560.0 & $+0.312$ & 0.085 \\
4929.0--4939.0 & $+0.412$ & 0.096 \\
5849.0--5859.0 & $+0.030$ & 0.226 \\
6135.0--6145.0 & $+0.272$ & 0.126 \\
\mbox{[La/Fe]} \\
4720.0--4750.0 & $+0.332$ & 0.261 \\
4915.0--4930.0 & $+0.210$ & 0.196 \\
4965.0--4990.0 & $+0.341$ & 0.336 \\
\hline
\end{longtable}
}

\onltab{2}{
\begin{longtable}{lcc}
\caption{\label{tab:iknabun}Individual abundance measurements for \object{IKN-5}.}\\ \hline \hline
Wavelength [\AA] & Value & Error \\ \hline
\endfirsthead
\caption{continued.}\\ \hline \hline
Wavelength [\AA] & Value & Error \\ \hline
\endhead
\hline
\endfoot
\mbox{[Fe/H]} \\
4600.0--4800.0 & $-1.868$ & 0.131 \\
4800.0--5000.0 & $-1.868$ & 0.106 \\
5000.0--5200.0 & $-2.066$ & 0.056 \\
5200.0--5400.0 & $-2.327$ & 0.070 \\
5400.0--5600.0 & $-2.376$ & 0.085 \\
5600.0--5800.0 & $-1.888$ & 0.101 \\
6000.0--6200.0 & $-1.967$ & 0.136 \\
\hline
\end{longtable}
}

\begin{table}
\begin{minipage}[t]{85mm}
\caption{Average integrated-light abundance measurements.}
\label{tab:abun}
\renewcommand{\footnoterule}{}
\begin{tabular}{lrlrl} \hline\hline
            & \multicolumn{2}{c}{\object{WLM GC}} & \multicolumn{2}{c}{\object{IKN-5}} \\
            & w.\ avg & rms$_w$ (N) & w.\ avg & rms$_w$ (N) 
            \\   \hline
$[$Fe/H$]$ & $-1.96$ & $0.08$ (9)   & $-2.11$ & 0.19 (7) \\
$[$Na/Fe$]$ & $+0.23$ & $0.15$ (1)  \\
$[$Mg/Fe$]$ & $+0.04$ & $0.15$ (5)  \\
$[$Ca/Fe$]$ & $+0.24$ & $0.11$ (6)  \\
$[$Sc/Fe$]$ & $+0.26$ & $0.20$ (6)  \\
$[$Ti/Fe$]$ & $+0.24$ & $0.08$ (9)  \\
$[$Cr/Fe$]$ & $-0.09$ & $0.15$ (5)  \\
$[$Mn/Fe$]$ & $-0.41$ & $0.11$ (2)  \\
$[$Y/Fe$]$ & $+0.01$ & $0.31$ (3)  \\
$[$Ba/Fe$]$ & $+0.32$ & $0.09$ (4)  \\
$[$La/Fe$]$ & $+0.27$ & $0.06$ (3)  \\
\hline
\end{tabular}
\end{minipage}
\tablefoot{
$N$ is the number of individual fits for each element (see Table~\ref{tab:wlmabun}--\ref{tab:iknabun}). For each entry we give the weighted average of the individual measurements and the weighted rms scatter.
}
\end{table}

For the final analysis, we adopted model HRDs based on isochrones with an age of 13 Gyr, $[\mathrm{\alpha/Fe}]=+0.2$ and the horizontal branch of the Galactic GC \object{NGC~6779} for both GCs. The fitting procedure discussed in Sect.~\ref{sec:analysis} converged towards input isochrone metallicities of $[\mathrm{Fe/H}]=-2.0$ for the \object{WLM GC} and $[\mathrm{Fe/H}]=-2.1$ for \object{IKN-5} (The grid of isochrones we used had a resolution of 0.1 dex in [Fe/H].).

The resulting individual abundance measurements are listed in Tables~\ref{tab:wlmabun} and \ref{tab:iknabun}. 
For each element, we list the wavelength bins where measurements were made, along with the best-fit abundance of the element in each bin and the associated formal 1-$\sigma$ uncertainty from the $\chi^2$ fit. For the WLM GC,
we mostly used the same wavelength bins as in the analysis of the Fornax GC spectra.
We have added extra bins to include the Mn {\sc i} triplet near 6020 \AA\ and the Na {\sc i} lines at 5683 \AA\ and 5688 \AA, as well as three bins sensitive to La. 
The weaker Mn lines around 6020 \AA\ yield abundances consistent (within 1.5 $\sigma$) with those around 4770 \AA\ used in our earlier work on Fornax but with larger errors.
For the initial overall scaling and the [Fe/H] measurements, we used 200 \AA\ bins for both IKN-5 and the WLM GC. Sample fits are shown in Fig.~\ref{fig:ca56}.

The final, average abundances are listed in Table~\ref{tab:abun}.
We give the weighted average of the individual measurements for each element, as well as the weighted rms, computed as
\begin{equation}
 \mathrm{rms}_w = \left(\frac{\sum w_i \left(\mathrm{[X/Fe]}_i - \langle \mathrm{[X/Fe]} \rangle\right)^2}{\sum w_i}\right)^{1/2}
\end{equation}
for weights
\begin{equation}
  w_i = 1/\sigma_i^2 .
\end{equation}
The $\sigma_i$ values are the uncertainties on the individual measurements from Table~\ref{tab:wlmabun} and \ref{tab:iknabun}. By applying these weights, the rms becomes less sensitive to outliers with large errors. 

The detailed analysis confirms the impression from Fig.~\ref{fig:fe5270} that both the WLM GC and IKN-5 are very metal-poor with metallicities near $\mathrm{[Fe/H]}\approx-2$. In this sense, they are similar to the four metal-poor GCs in the Fornax dSph and substantially more metal-poor than a typical GC in the Galactic halo.

\subsection{Sensitivity to input assumptions}
\label{sec:uncertainties}

\begin{table}
\begin{minipage}[t]{85mm}
\caption{Sensitivities to input assumptions for the \object{WLM GC}.}
\label{tab:abunvar}
\renewcommand{\footnoterule}{}
\begin{tabular}{lcccccc} \hline\hline
            & \multicolumn{2}{c}{$\Delta t$} & $\Delta$[Fe/H]$_i$ & $\Delta[\alpha/\mathrm{Fe}]$  & $\Delta_\mathrm{cont}$ \\ 
            & $-3$ Gyr & $-5$ Gyr & $+0.2$ dex & $+0.2$ dex & \\
             \hline
$\Delta$$[$Fe/H$]$ & $+0.013$ & $+0.086$ & $-0.046$ & $-0.093$ & $-0.001$ \\
$\Delta$$[$Na/Fe$]$ & $+0.004$ & $-0.036$ & $+0.018$ & $+0.037$ & $+0.131$ \\
$\Delta$$[$Mg/Fe$]$ & $+0.014$ & $+0.039$ & $+0.026$ & $+0.035$ & $-0.011$ \\
$\Delta$$[$Ca/Fe$]$ & $+0.005$ & $+0.013$ & $-0.005$ & $+0.018$ & $-0.005$ \\
$\Delta$$[$Sc/Fe$]$ & $+0.015$ & $-0.016$ & $+0.010$ & $-0.001$ & $-0.143$ \\
$\Delta$$[$Ti/Fe$]$ & $+0.008$ & $-0.001$ & $-0.009$ & $+0.004$ & $-0.026$ \\
$\Delta$$[$Cr/Fe$]$ & $-0.003$ & $+0.017$ & $-0.017$ & $-0.006$ & $+0.080$ \\
$\Delta$$[$Mn/Fe$]$ & $-0.006$ & $-0.030$ & $-0.012$ & $+0.008$ & $+0.151$ \\
$\Delta$$[$Y/Fe$]$ & $-0.006$ & $-0.049$ & $+0.043$ & $+0.018$ & $-0.188$ \\
$\Delta$$[$Ba/Fe$]$ & $+0.004$ & $+0.028$ & $-0.014$ & $-0.037$ & $-0.003$ \\
$\Delta$$[$La/Fe$]$ & $-0.034$ & $-0.050$ & $-0.009$ & $+0.004$ & $-0.209$ \\
\hline
\end{tabular}
\end{minipage}
\tablefoot{
This table lists the sensitivity of the derived abundance ratios to changes in the input isochrones of $\Delta t = -3$ Gyr and $-5$ Gyr, $\Delta$[Fe/H]$_i=+0.2$ dex, and $\Delta[\alpha$/Fe]=$+0.2$, which are relative to a reference age of 13 Gyr, an input metallicity of [Fe/H]$_i=-2.0$ and
[$\alpha$/Fe] = $+0.2$.
The last column lists the changes when adopting a modified procedure for the continuum scaling (see Sect.~\ref{sec:scal} for details).
}
\end{table}

The sensitivity of the integrated-light abundance analysis to various model assumptions was discussed extensively in \citetalias{Larsen2012a}. Uncertainties arising from the treatment of micro-turbulence and other aspects of the stellar atmosphere calculations, as well as those pertaining to the HRDs (luminosity functions, stochastic sampling of the IMF, and choice of CMD bins) were all found to affect the overall metallicity determinations by less than 0.1 dex. For most abundance ratios, the uncertainties due to these details were even smaller. One of the larger uncertainties was the extinction correction, through its effect on stellar temperatures as derived from the photometry. This is not so relevant here due to our use of theoretical isochrones, but we must now instead consider the uncertainties due to the choice of a particular isochrone. In the following, we discuss this and a couple of other sources of uncertainty.

\subsubsection{Input isochrone parameters}

We repeated the analysis of the WLM GC for other choices of ages and metallicities of the isochrones. The results are presented in Table~\ref{tab:abunvar}. Changing the age from 13 Gyr to 10 Gyr has only a very minor effect on most abundances and abundance ratios with typical changes of 0.01--0.02 dex. Even for an age as young as 8 Gyr, the metallicity changes by less than 0.1 dex. The changes are even smaller for most individual abundance ratios.
Note that this does not take the changes in horizontal branch morphology that would presumably occur at these younger ages into account. However, we do not consider the HB as a major uncertainty here, since its morphology is known from observations \citep{Hodge1999}.
Increasing the metallicity of the model isochrone by $+0.2$ dex causes the derived [Fe/H] to decrease by 0.05 dex, but most individual abundance ratios are only changed by 0.01--0.02 dex. 
We also looked at the effect of changing the [$\alpha$/Fe] ratio of the isochrones. We have assumed a moderate $\alpha$-enhancement of [$\alpha$/Fe]$=+0.2$ dex in our analysis but have repeated the analysis of the WLM GC for [$\alpha$/Fe]$=+0.4$ dex. The results are given in column 5 of Table~\ref{tab:abunvar}. The variations are generally small and comparable to those seen when varying the age or overall metallicity. 

\subsubsection{Scaling of model spectra}
\label{sec:scal}

A problem worth some consideration is the proper scaling of the continuum level. If the continuum is placed too low, spectral features appear too weak and abundances will be underestimated. As mentioned in Sect.~\ref{sec:analysis}, our ``continuum placement'' is done by applying an overall scaling to the spectra that may vary smoothly with wavelength. Since even our smaller fitting windows cover a significant fraction of a UVES echelle order ($\sim30$ \AA\ -- 50 \AA), we typically allow some curvature by using a second order polynomial for the scaling. For wider windows (e.g., those used for the Fe abundances), we use a third order spline function.
One potential concern is that the scaling may be affected by weak features that are present in the data but missing from the model spectra due to the inevitably incomplete line list \citep[e.g.,][]{Sakari2013}. The problem is worse at higher metallicities (where a larger fraction of the spectrum is noticeably affected by absorption features), and in \citetalias{Larsen2012a}, we found that our scaling procedure did indeed tend to place the continuum too low when analysing the spectrum of \object{Arcturus}. For \object{Arcturus}, this was addressed by identifying continuum regions in the \emph{observed} high S/N spectrum and by using only these regions for the scaling of the spectra. However, this procedure cannot be directly applied to our GC spectra, since their S/N is lower and the features weaker, so that a selection of ``continuum'' regions based on local maxima is more likely to select noise peaks (and thus introduce a bias in the opposite sense).

To assess the sensitivity of our analysis to the regions used for the scaling of the spectra, we adopted a slight modification of the procedure used in L12a. We used the \object{Arcturus} spectrum \citep{Hinkle2005} to identify continuum regions in a manner similar to what was done in \citetalias{Larsen2012a}. Since \object{Arcturus} has a much higher metallicity 
\citep[$\mathrm{[Fe/H]}\approx-0.5$;][]{Ramirez2011}
than the \object{WLM GC}, this should provide a very conservative estimate of regions free of significant absorption features in the WLM GC spectrum.
We first degraded the \object{Arcturus} spectrum to account for the velocity broadening of the WLM GC spectrum. We then identified ``continuum'' regions as pixels that had a flux $F > 0.95 \, F_\mathrm{max}$, where $F_\mathrm{max}$ is the maximum flux in a window of width $\pm5$ \AA\ around the pixel. This procedure eliminated a large fraction of the spectrum in the blue (leaving $\sim11$\% of the pixels in the interval 4400~\AA\ -- 4600~\AA), while more pixels were left at longer wavelengths (e.g., $\sim39$\% at 5250~\AA\ -- 5400~\AA).  We then flagged the corresponding regions in the WLM GC spectrum and used them for the overall scaling of the spectra. The remaining steps of the fitting procedure were, as far as possible, done in the same manner as for the original fits. However, due to the significantly reduced number of continuum pixels, we had to reduce the order of the polynomial fits used to match the model and observed spectra. 

The last column in Table~\ref{tab:abunvar} gives the changes in the abundances for the modified scaling procedure.
We see that the Fe abundance is virtually unaffected and the abundances of Mg, Ca, Ti and Ba also change very little. Moreover, the small changes in the abundances of these elements are \emph{negative} and, thus, in the opposite sense of what would be expected if the continuum level were set systematically too low in the original fits. Other elements show larger changes, which can be both positive and negative. We comment on Na in some detail, as this element is particularly important to the discussion later on. For this fit, about 52\% of the pixels were used for the scaling and the modified procedure yielded [Na/Fe]=$+0.36$ dex, while the original measurement was [Na/Fe]=$+0.23$ dex  (Table~\ref{tab:abun}). We note that the difference is within the $1 \sigma$ uncertainty on the original fit. 
The change in the abundance is, however, mostly due to the change of the scaling function from a second order to a first order polynomial, rather than the restrictions on the spectral regions used for the scaling: a fit \emph{without} any restrictions on the continuum pixels, using the first order scaling function, yielded [Na/Fe]=$+0.32$. Conversely, a second order scaling polynomial with Arcturus-based continuum pixels gave [Na/Fe]=$+0.25$.  Thus, in this case it also appears that the differences are not due to a systematic error in the continuum scaling but due to the inherent uncertainty in determining the proper scaling of the model spectra for noisy data. We have found no obvious reason to prefer one set of fits over the other; for example, the rms$_w$ increases for the modified fits in some cases (Mg, Ca, Ti, Cr, Ba, and La) and decreases in others (Sc, Mn, and Y). We therefore use the ``standard'' fits in the remainder of this paper for consistency with \citetalias{Larsen2012a}.

\subsubsection{Empirical CMDs vs.\ model isochrones}

A final point concerns possible systematic differences between the semi-empirical approach used in our analysis of the Fornax GCs and the isochrone-based approach used here. To investigate this, we repeated the analysis of the Fornax 5 cluster with an isochrone-based HRD. Rather than remeasuring the full set of elements, we restricted this test to Fe, Mg, and Ca. Based on the other tests we have carried out, the systematic uncertainties on the abundances of Mg and Ca appear to be quite typical for the full set of elements. We used the same model HRD employed for IKN-5, where $\mathrm{[Fe/H]}=-2.1$ and $t=13$ Gyr, and the exact same wavelength bins used in the original analysis of Fornax 5 \citepalias{Larsen2012a}. In this way, we found a metallicity of $\mathrm{[Fe/H]}=-2.08$ and abundance ratios of $\mathrm{[Mg/Fe]}=+0.11$ and $\mathrm{[Ca/Fe]} = +0.24$ for Fornax 5. 
This may be compared with the values of $\mathrm{[Fe/H]}=-2.09$, $\mathrm{[Mg/Fe]}=+0.13$, and $\mathrm{[Ca/Fe]}=+0.27$ found in \citetalias{Larsen2012a} from the analysis based on the empirical CMD. This confirms our conclusion from Sect.~\ref{sec:wlmgc} that the switch to theoretical HRDs does not introduce major systematic differences with respect to the analysis based on empirical CMDs. 

\section{Discussion}

\subsection{Overall metallicities and implications for the formation of globular clusters}

With our new analysis of the WLM GC and IKN-5 combined with the existing data for the Fornax GCs, we now have accurate metallicity determinations for GCs in three dwarf galaxies. We begin by discussing the implications of these measurements for GC-formation scenarios.

In the \object{Fornax dSph}, \citetalias{Larsen2012} found that about 20\%--25\% of the metal-poor stars (with $\mathrm{[Fe/H]}<-2$) belong to the four metal-poor GCs. Let us now examine how this compares with the other dwarfs. From Fig.~2 of \citet{Lianou2010}, about 7\% of the RGB stars in the IKN dSph have $\mathrm{[Fe/H]}<-2$. This is quite similar to the corresponding fraction in Fornax but with an integrated magnitude of $M_V\approx-11.5$ \citep{Georgiev2009} \object{IKN} is much fainter than Fornax. If we simply scale by number, the integrated magnitude of the metal-poor stars is then $M_V\approx-8.6$. We do not know the metallicities of most of the GCs in IKN, but IKN-5 by itself has $M_V=-8.5$ \citep{Georgiev2009}. In other words, \emph{the luminosity of this single GC appears to be roughly similar to that of all the metal-poor field stars in the whole galaxy combined}. Of course, the metallicity distribution of the stars in IKN is less well constrained than in Fornax, and the metallicity of IKN-5 is somewhat uncertain, but even if we count \emph{all} stars with $\mathrm{[Fe/H]}<-1.5$ according to \citet{Lianou2010}, the GC:field ratio is still 1:4, which is similar to Fornax. 
This must be considered a conservative estimate, since it is already clear from Fig.~\ref{fig:fe5270} that IKN-5 has a metallicity well below $\mathrm{[Fe/H]}=-1.5$. Furthermore, the proper comparison (in terms of constraining mass loss from the GC) would presumably be with field stars in a narrow metallicity range around that of IKN-5, since GCs typically do not display any significant internal metallicity spread.

\begin{figure}
\centering
\includegraphics[width=\columnwidth]{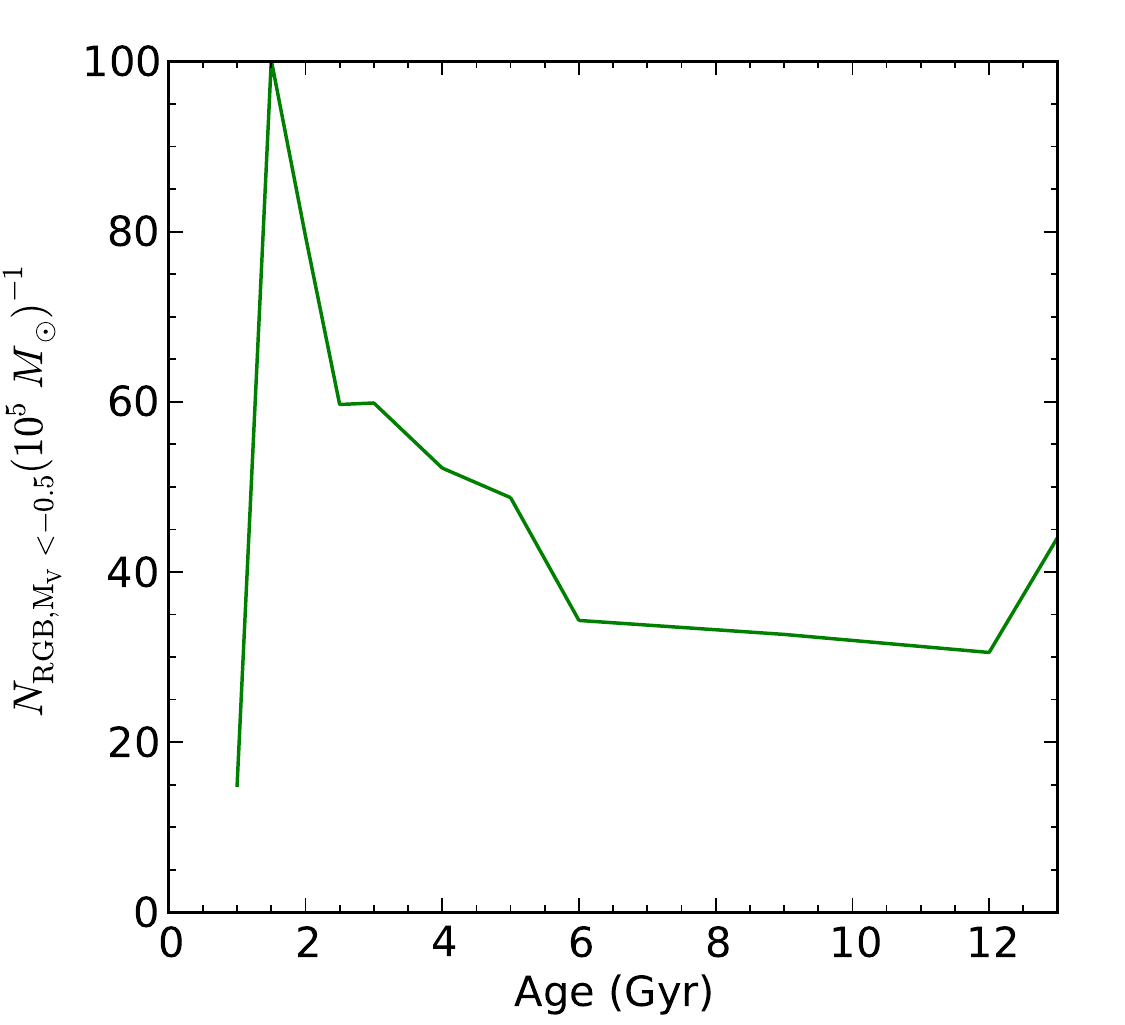}
\caption{\label{fig:nrgb}Number of RGB stars per $10^5$ M$_\odot$ as a function of age.}
\end{figure}

In the case of WLM, a large fraction of the stars are relatively young \citep{Dolphin2000} so that the integrated luminosities of WLM itself and the GC cannot be directly compared. From a multi-band integrated-light spectral energy distribution fitting, \citet{Zhang2012} found a total stellar mass of $\sim1.6\times10^7 M_\odot$ for the galaxy. If we use the relations in \citet{Bell2001} to estimate the mass-to-light ratio instead, then we find $(M/L)_V = 0.71$ for the observed integrated colour, $B-V=0.42$ \citep[RC3;][]{DeVaucouleurs1991}. The Bell \& de Jong relations assume a Salpeter IMF with lower and upper mass limits of 0.1 M$_\odot$ and 125 M$_\odot$ respectively, but with $M/L$ ratios scaled down by 30\%.
For an apparent magnitude $V=10.59$ (RC3) and $A_V=0.104$, we get $M_V=-14.44$ (i.e., $L_V = 5.1\times10^7 L_{\odot, V}$), which yields a significantly greater mass of $M = 3.6\times10^7 M_\odot$ compared to the \citet{Zhang2012} estimate. If we use the simple stellar population models of \citet{Bruzual2003} for a \citet{Chabrier2003} IMF and assume a constant SFR for $t<13$ Gyr, then we find $(M/L)_V = 0.73$ and $B-V=0.41$ (for $Z=0.004$), which closely agrees with the observed colour and the Bell \& de Jong $(M/L)_V$ ratio. For a lower metallicity of $Z=0.0004$ we get $(M/L)_V=0.68$ and $B-V=0.35$, which is a similar $(M/L)_V$ but a poorer match to the observed colour. We thus consider $[1.6-3.6]\times10^7 M_\odot$ a likely range for the total stellar mass of the WLM galaxy.

The metallicity distribution of RGB stars in WLM has been studied spectroscopically by \citet{Leaman2013}, who found a mean of $\mathrm{[Fe/H]} = -1.28$, which is again substantially higher than for the GC. 
Interestingly, our metallicity estimate for the WLM GC closely agrees with the age-metallicity relation for stars in WLM found by \citet{Leaman2013}.  
From their Figure~5, about 8\% of the RGB stars have $\mathrm{[Fe/H]}\leq-2$.
To convert this number fraction into a mass fraction, we need to take the RGB lifetimes and the rate at which RGB stars are produced as a function of age and metallicity into account \citep[e.g.][]{Renzini1986}. Taking this information from the \citet{Dotter2007} isochrones and assuming a Salpeter IMF with a lower mass limit of 0.15 M$_\odot$, we obtain Fig.~\ref{fig:nrgb}, which shows the number of RGB stars brighter than $M_V=-0.5$ per $10^5$ M$_\odot$ as a function of age. At each age, we have used isochrones according to the age-metallicity relation of \citet{Leaman2013}, and integrated the IMF over the mass range that corresponds to the initial masses of RGB stars between $M_V=-0.5$ and the tip of the RGB.
If we assume that the metal-poor stars formed between 12 and 13 Gyr ago and the rest of the stars were formed at a constant rate from the present until 12 Gyr ago, then we find that a metal-poor RGB \emph{number} fraction of 8\% corresponds to a \emph{mass} fraction of 8.7\%.
The actual SFH of WLM is poorly known, since even HST imaging can only probe main sequence stars younger than about 2--3 Gyr \citep{Dolphin2000}. \citet{Weisz2008} estimated that about 10\%, 44\%, and 46\% of the stars formed 0--1 Gyr, 1--6 Gyr, and more than 6 Gyr ago, respectively. This is not very different from the fractions corresponding to a constant SFH, and integrating the curve in Fig.~\ref{fig:nrgb} over this SFH only changes the mass fraction by about 0.1\%. It thus appears that the metal-poor vs. metal-rich RGB number fraction is quite representative of the corresponding mass fraction.

For an age of 13 Gyr, the \citet{Bruzual2003} models predict $(M/L)_V = 1.94$ or 1.90 for $Z=0.0001$ or $Z=0.0004$, respectively. This is similar to the typical $(M/L)_V \approx 2$ measured for metal-poor GCs in M31 by \citet{Strader2009}. For an integrated magnitude of $M_V=-8.96$, the WLM GC then has a mass of $M \sim 6.3\times10^5 M_\odot$. 
Scaling total galaxy masses of $1.6\times10^7 M_\odot$ or $3.6\times10^7 M_\odot$ by 8.7\%, the GC then accounts for 31\% or 17\% of the metal-poor stellar mass in WLM.
As a consistency check of the mass of the WLM GC, we can use our estimate of the velocity dispersion of $\sigma_\mathrm{1D}= 9.6$ km/s (Sect.~\ref{sec:wlmgc}) combined with the half-light radius of 3.5 pc \citep{Stephens2006}.  The virial mass \citep[e.g.,][]{Larsen2002b} becomes $M\sim7.5\times10^5 M_\odot$, which is in good agreement with the photometric estimate.

It appears that the high GC-field star ratio in the Fornax dSph is neither unique nor particularly extreme. 
The high GC-to-field star ratios in dwarf galaxies  have important implications for understanding star cluster formation and evolution, as they constrain the amount of mass that could have been lost from clusters to the field \citepalias{Larsen2012}.
To illustrate this, consider a population of star clusters that formed with a total initial mass $M_\mathrm{init}$ and is observed at a later time when it has lost a fraction $f_\mathrm{lost}$ of this initial mass. Then the observed mass in clusters is clearly  $M_\mathrm{GC} = M_\mathrm{init} \times (1 - f_\mathrm{lost})$. If no stars have been lost from the system and all stars were in clusters initially, then $M_\mathrm{init} = M_\mathrm{GC} + M_\mathrm{field}$ where $M_\mathrm{field} = f_\mathrm{lost}\times M_\mathrm{init}$. Here, we have ignored that some mass is lost due to stellar evolution; to first order, this affects the field and GCs equally. Hence, $1 - f_\mathrm{lost} = M_\mathrm{GC}/M_\mathrm{init} = M_\mathrm{GC}/(M_\mathrm{GC} + M_\mathrm{field})$. Then, if  $M_\mathrm{GC}/(M_\mathrm{GC} + M_\mathrm{field}) = 17\%-31\%$, we have $f_\mathrm{lost} = 69\%-83\%$. This is the \emph{maximum} fraction of mass that could have been lost from GCs to the field, regardless of whether individual clusters lost part of their mass, or whether some clusters disrupted completely. If any stars were formed as ``field'' stars, then our estimate of $f_\mathrm{lost}$ is an \emph{upper} limit. 
Similarly, if the field consists partly of fully disrupted clusters, then the remaining clusters must have lost less than $f_\mathrm{lost}$ of their mass.
An important caveat is that field stars may have been lost preferentially from the system (e.g., via tidal stripping), in which case $f_\mathrm{lost}$ could have been higher. In the Fornax dSph, $N$-body simulations suggest that no significant amount of stars have been lost \citep{Penarrubia2009}. This is less well constrained in WLM and IKN, although WLM is one of the most isolated galaxies in the Local Group, possibly near the zero-velocity surface, and may not have interacted significantly with other galaxies during its lifetime \citep{Sandage1985,Minniti1996a}.

\citet{DAntona2013} have argued that it may be possible to accommodate the high $M_\mathrm{GC}/M_\mathrm{field}$ ratio observed in Fornax, even if the polluters of the second generation stars in GCs are massive AGB stars. However, this requires some fine-tuning of the second generation initial mass function, so as to avoid formation of high-mass stars and thereby reduce the mass-budget problem. It also leaves little room for formation of now-disrupted, low-mass clusters or bona-fide field stars in the same metallicity range as the surviving GCs.
\citet{Bastian2013a} have proposed that the abundance anomalies observed in GCs may result from the accretion of ejecta from massive interacting binaries onto proto-stellar discs of low-mass stars. The mass-budget problem in this scenario is far less severe, although it requires that a large fraction of the total mass in stars with $M > 10 \, M_\odot$ is accreted onto low-mass stars. 
Another proposed scenario is that the proto-cluster gas was reprocessed and polluted by super-massive stars formed in run-away collisions during the early stages of the cluster formation \citep{Denissenkov2013}.

\subsection{Detailed chemical composition of GCs in dwarf galaxies}

\begin{figure*}
\centering
\includegraphics[width=\textwidth]{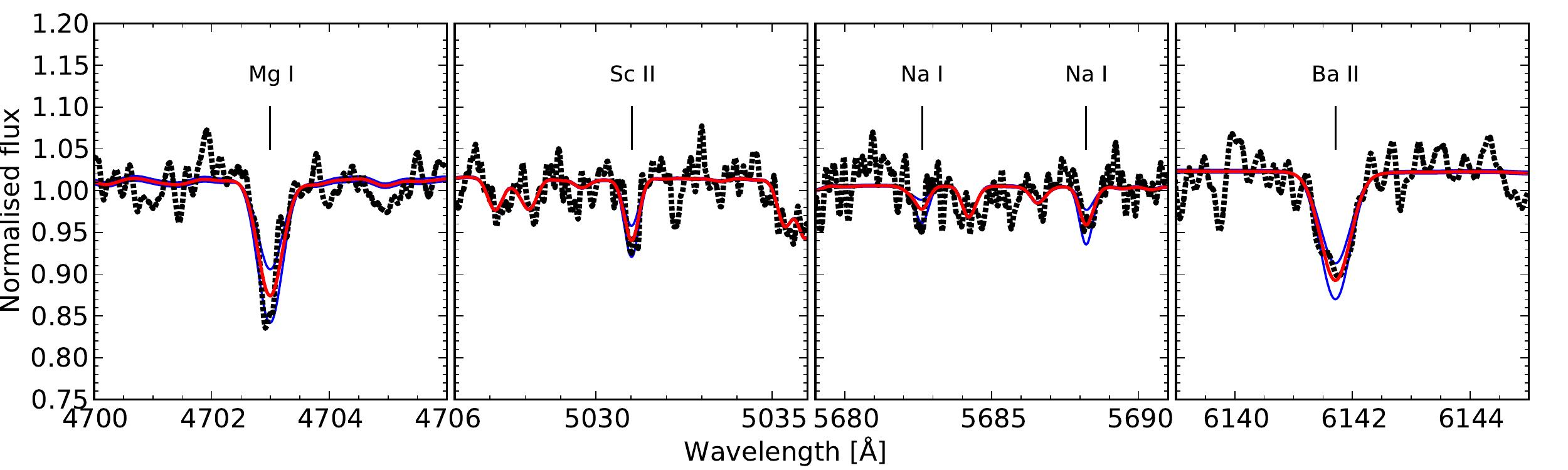}
\caption{\label{fig:mult}Example fits to individual features. The black dotted curves show the observed spectrum of the WLM GC, while the thick solid curves (red in the on-line version) are the best-fitting model spectra for the wavelength bins shown in the figure. The thin curves (blue in the on-line edition) show models where the abundances of the corresponding elements have been varied by $\pm0.3$ dex.
The spectra have been smoothed by a Gaussian kernel with $\sigma = 1.5$ pixels.}
\end{figure*}

We now turn to the detailed chemical composition of the WLM GC. A few sample fits to individual spectral features are shown in Fig.~\ref{fig:mult}.  We also show models where the abundances have been varied by $\pm0.3$ dex.
This analysis addresses two main issues: first, the scenarios for self-enrichment and multiple stellar populations within GCs, as discussed above, were developed to explain
the anomalous abundances of several of the light elements in Galactic GCs and the evidence from colour-magnitude diagrams \citep[e.g.,][]{Gratton2012}. 
It is therefore important to establish how similar the extragalactic GCs, particularly those in dwarf galaxies, are to Galactic GCs in this respect. Second, we can take advantage of the observation that elements heavier than Al are generally not affected by these anomalies, such that their abundances may be expected to reflect the general chemical enrichment histories of their parent galaxies up to the time when the GCs formed.

\subsubsection{Light elements: Na and Mg}
\label{sec:light}

\begin{figure}
\centering
\includegraphics[width=85mm]{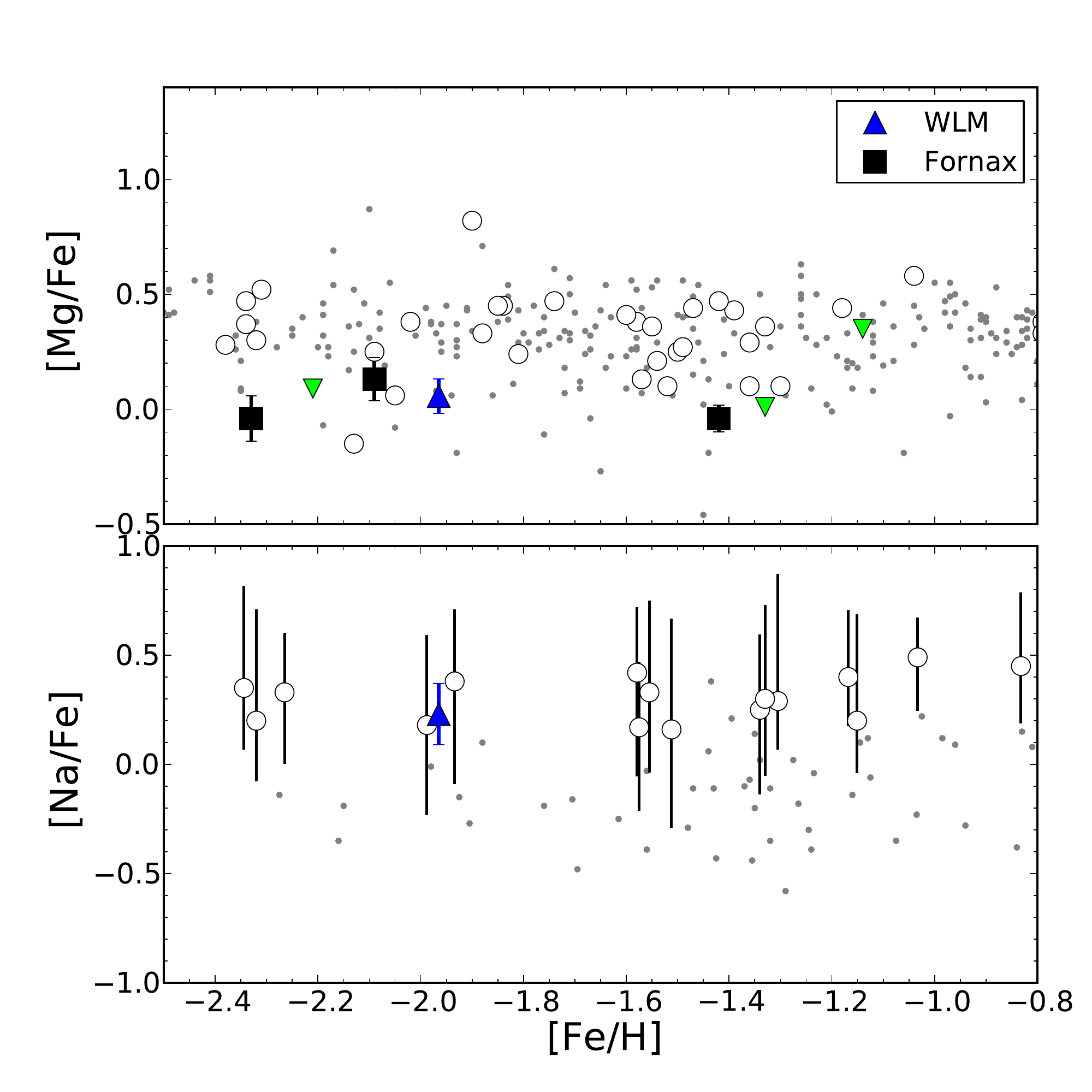}
\caption{\label{fig:light}Abundances of Na and Mg. Large symbols with error bars are data from this paper and \citetalias{Larsen2012a}, as indicated in the legend.  Upside-down triangles (green in the on-line version) are data for M31 GCs \citep{Colucci2009}. 
In the top panel, open circles are mean abundances for Milky Way GCs from \citet{Pritzl2005}. In the lower panel, 
 open circles and the vertical lines represent the mean abundances and their range for Milky Way GCs from \citet{Carretta2009}. 
Small grey dots are individual Milky Way stars (\citealt{Venn2004} for Mg and \citealt{Ishigaki2013} for Na.)}
\end{figure}

Figure~\ref{fig:light} shows our measurements of  [Na/Fe] and [Mg/Fe] for the WLM GC. For Mg, we have also included data for the Fornax GCs from \citetalias{Larsen2012a}. 
The Na 5683/5688 \AA\ lines in the Fornax GC spectra are, unfortunately, affected by a detector blemish that prevented measurement of Na for these clusters.
In this and the following figures, 
the error bars in the $y$-direction are computed as $\sigma = \mathrm{rms}_w/\sqrt{N-1}$, where $N$ is the number of individual measurements. For $N=1$, we use the formal errors on the fits from Table~\ref{tab:wlmabun}. For clarity, we omit the error bars in the $x$-direction.
Also included in Fig.~\ref{fig:light} are data for individual field stars in the Milky Way \citep{Venn2004,Ishigaki2013}, for stars in Milky Way GCs \citep{Pritzl2005,Carretta2009}, and for integrated-light measurements for GCs in M31 \citep{Colucci2009}.  

Perhaps, the best-known and most common of the chemical anomalies in Milky Way GCs is the Na-O anti-correlation, whereby a large fraction of the cluster stars have high [Na/Fe] and low [O/Fe] ratios compared to field stars. Some GCs also display a Mg-Al anti-correlation with depleted [Mg/Fe] and elevated [Al/Fe] for a fraction of the stars \citep{Gratton2001,Carretta2009}.  We might expect that integrated-light observations of such clusters yield  ``average''  abundances of the corresponding elements with enhanced [Na/Fe] and [Al/Fe] ratios and depleted [O/Fe] and [Mg/Fe] ratios compared to the canonical abundance patterns of field stars.
Unfortunately, neither O nor Al has suitable lines within the wavelength range covered by our data, so we cannot directly establish whether we are actually detecting these anti-correlations in integrated light. 

Nevertheless, the [Na/Fe] ratio in the WLM GC is clearly higher than for Milky Way field stars but quite similar to the mean [Na/Fe] ratios of Milky Way GCs. We do not have information about the Na abundances of individual stars in WLM, but other dwarfs appear similar to the Milky Way in terms of the [Na/Fe] abundances at these low metallicities \citep{Tolstoy2009}.
We note that the \citet{Carretta2009} Na abundances for Milky Way GCs have been adjusted for non-LTE effects, while no such corrections have been applied to our WLM GC data or the individual Milky Way  stars \citep{Ishigaki2013}. Doing so for the integrated-light analysis is not straight-forward, as these corrections depend on metallicity, $T_\mathrm{eff}$ and $\log g$ and must be applied individually to each star or cmd-bin. For cool giants, the NLTE corrections can increase the Na abundance by $0.2-0.3$ dex, although the corrections decrease with increasing temperature and surface gravity and even become negative for dwarf stars \citep{Gratton1999,Takeda2003}.  \citet{Ishigaki2013} estimate that NLTE corrections are typically negative but amount to less than about 0.1 dex for their sample of Milky Way stars and the lines used in their analysis. In summary, NLTE corrections would tend to shift the data points for individual stars (small dots) in Fig.~\ref{fig:light} downwards (by $\sim0.1$ dex or less) and the WLM GC upwards (by no more than $\sim0.2$ dex). 
The observed offsets between GC and field star data would thus be preserved and possibly even be enhanced.
We also note that changes in the definition of spectral regions used to match the scaling of our model and observed spectra tend to increase the [Na/Fe] ratio for the WLM GC (Sect.~\ref{sec:uncertainties}).

As noted previously \citep{Colucci2009,Larsen2012a}, the integrated-light [Mg/Fe] ratios tend to be lower for the GCs than for the Milky Way field stars. This is, at least qualitatively, also consistent with the presence of a Mg-Al anticorrelation for individual stars.

A particularly well-studied case in the Milky Way is the globular cluster \object{M13}, which shows clear Na-O and Mg-Al anti-correlations with a spread in the [Mg/Fe] ratios from $-0.2$ to $+0.4$ and in [Na/Fe] from $-0.3$ to $+0.6$. The ratios obtained by averaging over the individual star measurements are $\langle[\mbox{Mg/Fe}]\rangle = +0.11$ and $\langle[\mbox{Na/Fe}]\rangle = +0.21$ \citep{Sneden2004}, which is a combination rather similar to the one we find in the WLM GC.
From integrated-light analysis, \citet{Sakari2013} found $\mathrm{[Mg/Fe]}=+0.14\pm0.10$ and $\mathrm{[Na/Fe]}=+0.33\pm0.16$ for \object{M13}, which agrees with the average of the individual stellar measurements. This suggests that we might indeed be seeing these anti-correlations in the integrated-light \object{WLM GC} spectrum.

The GC \object{M13} may, however, not be a typical case in the Milky Way. In most Galactic GCs studied in detail to date, the spread in [Mg/Fe] is substantially smaller than in \object{M13}, and [Mg/Fe] ratios as low as those seen in our integrated-light measurements \citep[and those of][]{Colucci2009} are relatively uncommon. This is also evident from Fig.~\ref{fig:light}.
The \citet{Pritzl2005} compilation lists 41 Milky Way GCs with [Mg/Fe] measurements, of which only nine have an average $\mathrm{[Mg/Fe]} < 0.2$. Of the 18 GCs in the \citet{Carretta2009} sample, only one (\object{NGC~2808}) shows a significant number of stars with $\mathrm{[Mg/Fe]}<0.2-0.3$  (M13 was not included in that study.). 
It thus remains puzzling that strongly depleted [Mg/Fe] ratios seem to appear so commonly in integrated-light analyses of extragalactic GCs. 

It is, of course, not obvious that the integrated-light analysis should yield exactly the same result as the straight average of the logarithmic abundances of individual stars. For weak lines, the integrated-light analysis may be expected to give abundances roughly similar to the average of the individual abundances on a \emph{linear} scale, at least to the extent that the stars with different abundances are otherwise similar.
More work would be desirable to better understand the relation between the light element abundances of individual stars in GCs and those measured from integrated light.

\subsubsection{$\alpha$-elements: Ca and Ti}

\begin{figure}
\centering
\includegraphics[width=85mm]{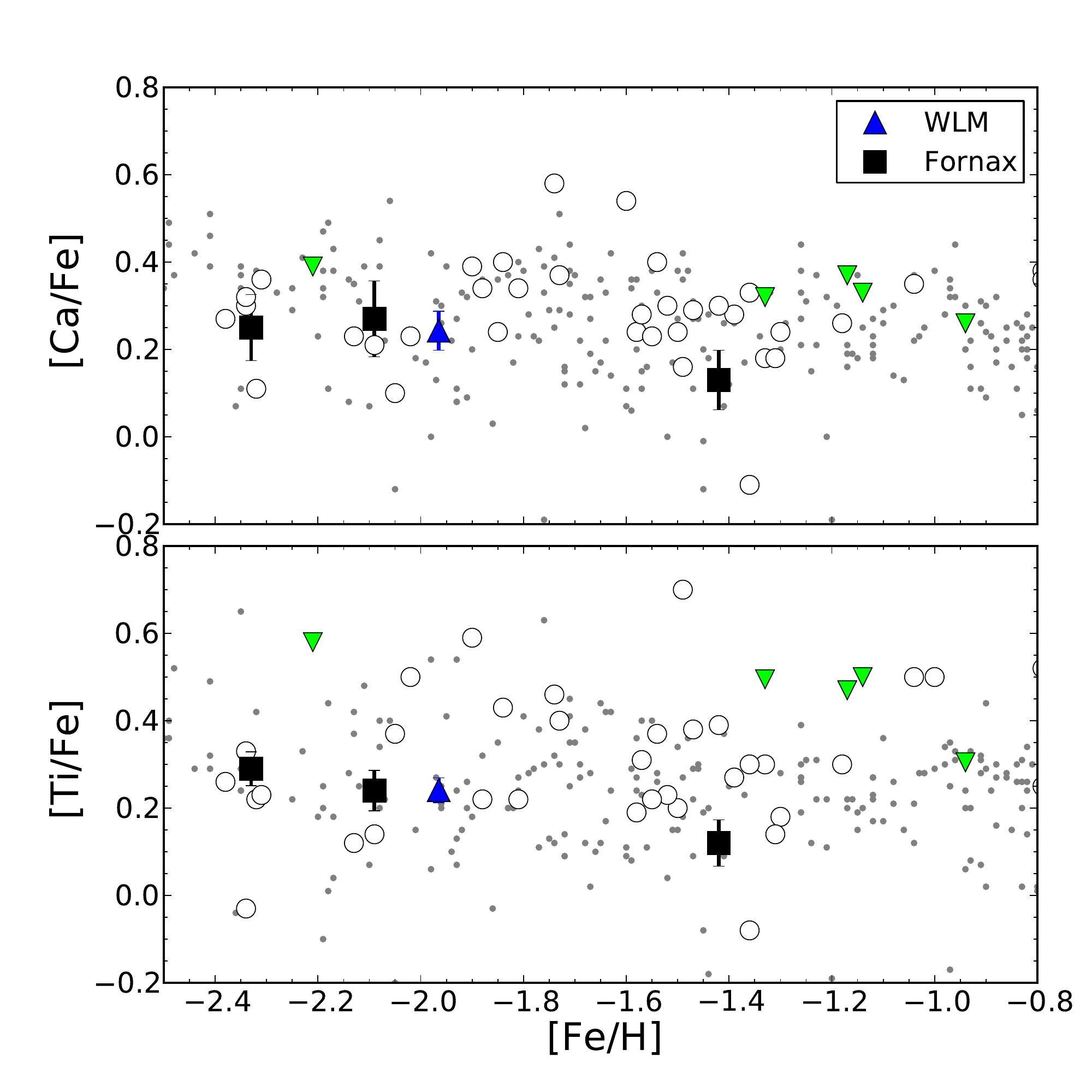}
\caption{\label{fig:alphafe}$\alpha$-element abundances. Large symbols with error bars are data from this paper and \citetalias{Larsen2012a}, as indicated in the legend.
The open circles are data for Milky Way GCs \citep{Pritzl2005}.  Upside-down triangles are data for M31 GCs \citep{Colucci2009}. Small grey dots are individual Milky Way stars \citep{Venn2004}.}
\end{figure}

In Fig.~\ref{fig:alphafe}, we compare the [Ca/Fe] and [Ti/Fe] abundance ratios with the literature data.  
The [Ca/Fe] and [Ti/Fe]  ratios for the GCs in the WLM and Fornax dwarfs are fairly similar to those in Milky Way GCs and individual stars at the corresponding metallicities. The most significant exception is Fornax 4, but the lower [Ca/Fe] and [Ti/Fe] ratios in this cluster are actually consistent with the trend seen for field stars in dwarfs at these metallicities \citep{Tolstoy2009}. The ``normal'' enhanced [Ca/Fe] and [Ti/Fe] ratios suggest that the low [Mg/Fe] ratios discussed in Sect.~\ref{sec:light} are not simply due to an overall depletion of the $\alpha$-element abundances.

\subsubsection{Iron-peak elements: Cr, Sc, and Mn}

\begin{figure}
\centering
\includegraphics[width=85mm]{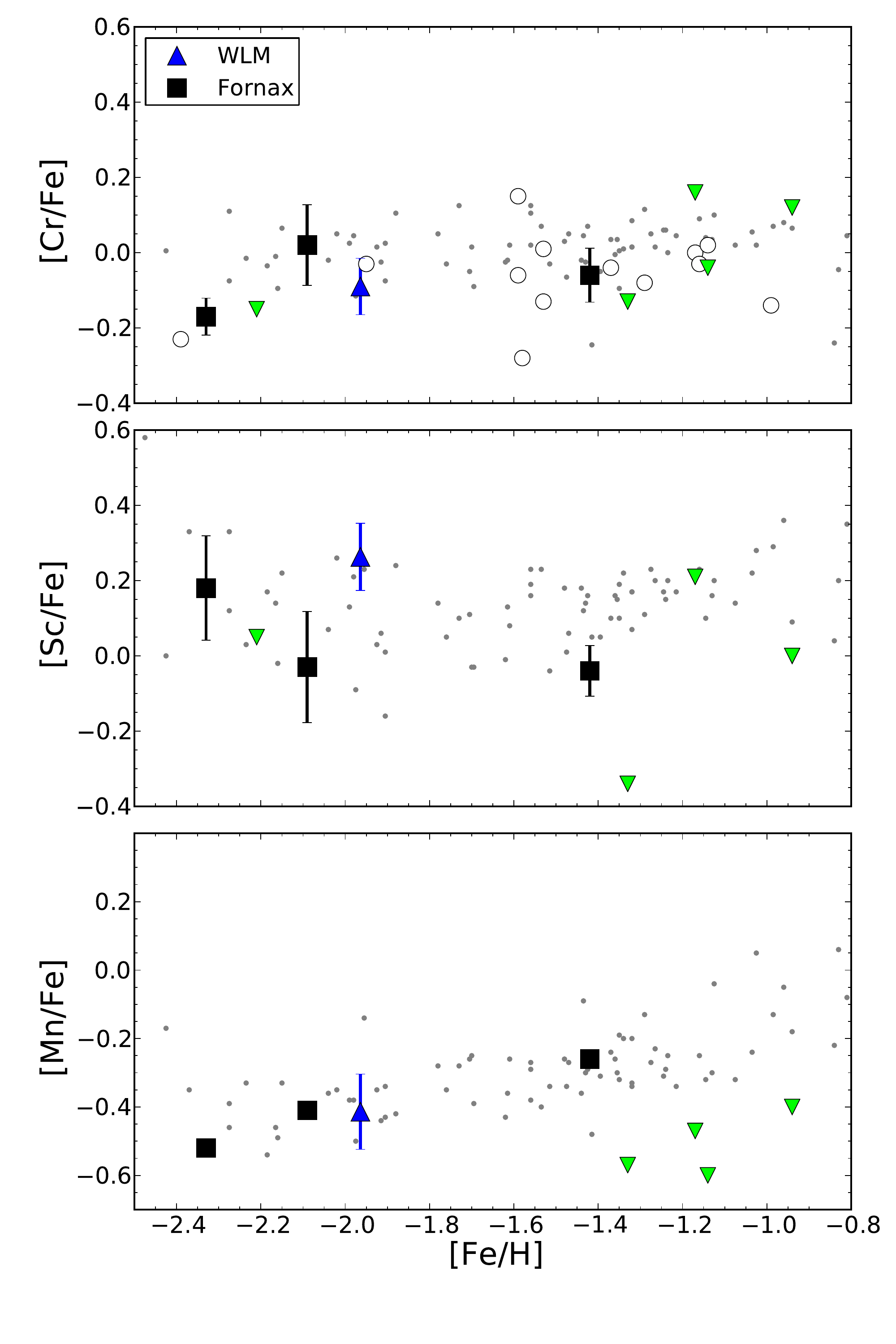}
\caption{\label{fig:fepeak}Iron-peak element abundances. Symbols are the same as in Fig.~\ref{fig:alphafe}, except that the Milky Way stellar data (small grey dots) are from \citet{Ishigaki2013} and Milky Way GC data are from \citet{Roediger2014}.}
\end{figure}

Our measurements of the iron-peak elements Cr, Sc, and Mn are shown in Fig.~\ref{fig:fepeak}. All three elements generally follow the trends seen in Milky Way stars \citep{Ishigaki2013}.
The \citet{Pritzl2005} compilation does not include these elements, but we have included the [Cr/Fe] data from the compilation in \citet{Roediger2014}.
For Sc and Mn, a direct comparison with Milky Way GCs is unfortunately not possible, but studies of individual Milky Way GCs tend to find abundances of these elements similar to those of field stars \citep[e.g.,][]{Carretta2006,Caliskan2012}.

We find [Cr/Fe] ratios close to zero with little correlation with [Fe/H], which is similar to results for the Milky Way and the LMC \citep{Gratton1991,Johnson2006a}. The modifications to the fitting procedure discussed in Sect.~\ref{sec:scal} would change [Cr/Fe] by $+0.08$ dex for the WLM GC, improving the agreement with the Milky Way data. There may be a tendency for the most metal-poor clusters ($\mathrm{[Fe/H]}\la-2$) to have slightly negative [Cr/Fe] ratios. The scatter in the [Cr/Fe] ratios is small, comparable to the errors.

The [Sc/Fe] ratio is solar or slightly super-solar over the whole metallicity range probed here, while Mn differs from the other Fe-peak elements in deviating more from Solar-scaled abundances. The [Mn/Fe] ratios of Milky Way stars increase from $\mathrm{[Mn/Fe]}\approx-0.5$ at $\mathrm{[Fe/H]}<-2$ to roughly solar at $\mathrm{[Fe/H]}>-1$. In general, our integrated-light measurements for the GCs in WLM and Fornax follow this trend. 
The modified fits would shift [Sc/Fe]  by $-0.14$ dex and [Mn/Fe] by $+0.15$ dex for the WLM GC, maintaining good agreement with the Milky Way data in both cases.
However, the four M31 GCs \citep{Colucci2009} fall well below the Milky Way data, perhaps suggesting different trends for the Milky Way and (some) extragalactic GCs at higher metallicities. 

The spectra of both Mn and Sc exhibit hyperfine structure. This can have a significant effect on abundance measurements, although mostly at metallicities greater than $\mathrm{[Fe/H]}\approx-1$ \citep{Prochaska2000}. 
Hyperfine splitting is only included for a few Mn {\sc i} and Sc {\sc ii} lines in the Castelli line list, but we tested our results by remeasuring the Mn and Sc abundances using the more complete line list now available from the Kurucz web site. The $gf$ values are the same for these elements in the two line lists, but the new Kurucz list includes hyperfine structure for a large number of Mn and Sc lines. The resulting changes in the abundances were very small, $\Delta [\mathrm{Mn/Fe}] = -0.02$ dex and $\Delta [\mathrm{Sc/Fe}] = - 0.01$ dex.

\subsubsection{Neutron-capture elements: Y, Ba, and La}

\begin{figure}
\centering
\includegraphics[width=85mm]{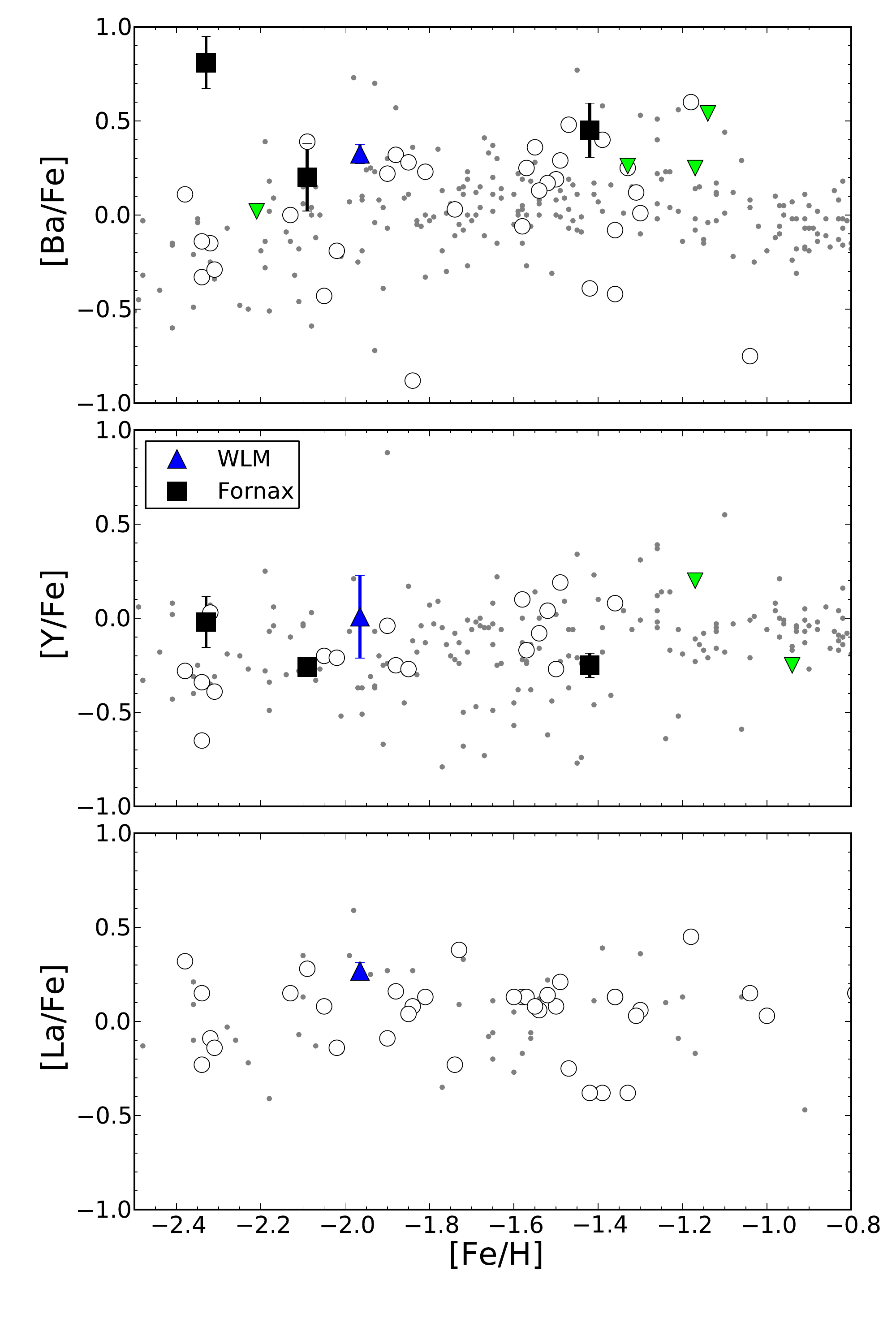}
\caption{\label{fig:ncap}Neutron-capture element abundances. Symbols are the same as in Fig.~\ref{fig:alphafe}.}
\end{figure}

Figure~\ref{fig:ncap} shows our measurements of the three neutron-capture elements, Y, Ba, and La.  
Apart from the most metal-poor of the Fornax GCs (\object{Fornax 3}), the abundance ratios of all three elements fall within the ranges covered by metal-poor Milky Way GCs and halo stars, although the Ba abundances tend to be somewhat higher than typical Milky Way GC values.
The Ba lines are generally well fit by our model spectra and blending does not appear to be a major issue for these lines. However, the Ba abundances are sensitive to the relative fractions of the various Ba isotopes, since hyperfine splitting desaturates the lines for  $^{135}$Ba and $^{137}$Ba, while $^{138}$Ba has zero nuclear spin and, thus, no hyperfine structure \citep{Rutten1978}. The line list includes hyperfine splitting for the Ba lines, but  the \texttt{SYNTHE} code assumes the Solar system isotope ratios, which are dominated by $^{138}$Ba \citep{Anders1989}. The $r$-process dominated isotope ratios of \citet{McWilliam1998} have a higher fraction of $^{135}$Ba and $^{137}$Ba, so that hyperfine structure is more important and a given line strength corresponds to lower abundances. If we adopted the isotope ratios of \citet{McWilliam1998}, then the Ba abundances decreased by about 0.10 dex.  
Another uncertainty is the oscillator strength of the Ba {\sc ii} 4934 \AA\ line. \citet{McWilliam1998} gives this as $\log gf = -0.15$, essentially the same value listed in the NIST database \citep{NIST}, and this is also the value used in the most recent Kurucz line list. In contrast, the Castelli list has $\log gf = -0.45$. The Ba abundance inferred from this line would thus decrease by 0.30 dex if we used the higher $\log gf$ value. Nevertheless, since the overall Ba abundance is an average from four lines, the overall effect is less than 0.10 dex.
It is thus likely that our [Ba/Fe] ratios should be adjusted downwards by 0.1--0.2 dex, which would bring most of them into closer agreement with typical Milky Way abundances (although Fornax 3 would remain an outlier).

The [La/Fe] ratio for the WLM GC is quite uncertain because most of the La features are relatively weak and/or blended with other lines. Furthermore,  La  has significant hyperfine structure \citep{Ivans2006}, which is not included in the Kurucz/Castelli line list. Nevertheless, the three individual fits give quite consistent results, and we have chosen to include the La measurement here for comparison with Ba, since the two elements are expected to share a common nucleosynthetic history: they are neighbours in the periodic table and both belong to the second $s$-process peak. Indeed, the [La/Fe]  and [Ba/Fe] ratios observed in the WLM GC behave very similarly with respect to the Milky Way data.

\subsection{GCs and their parent galaxies}

\begin{figure}
\centering
\includegraphics[width=85mm]{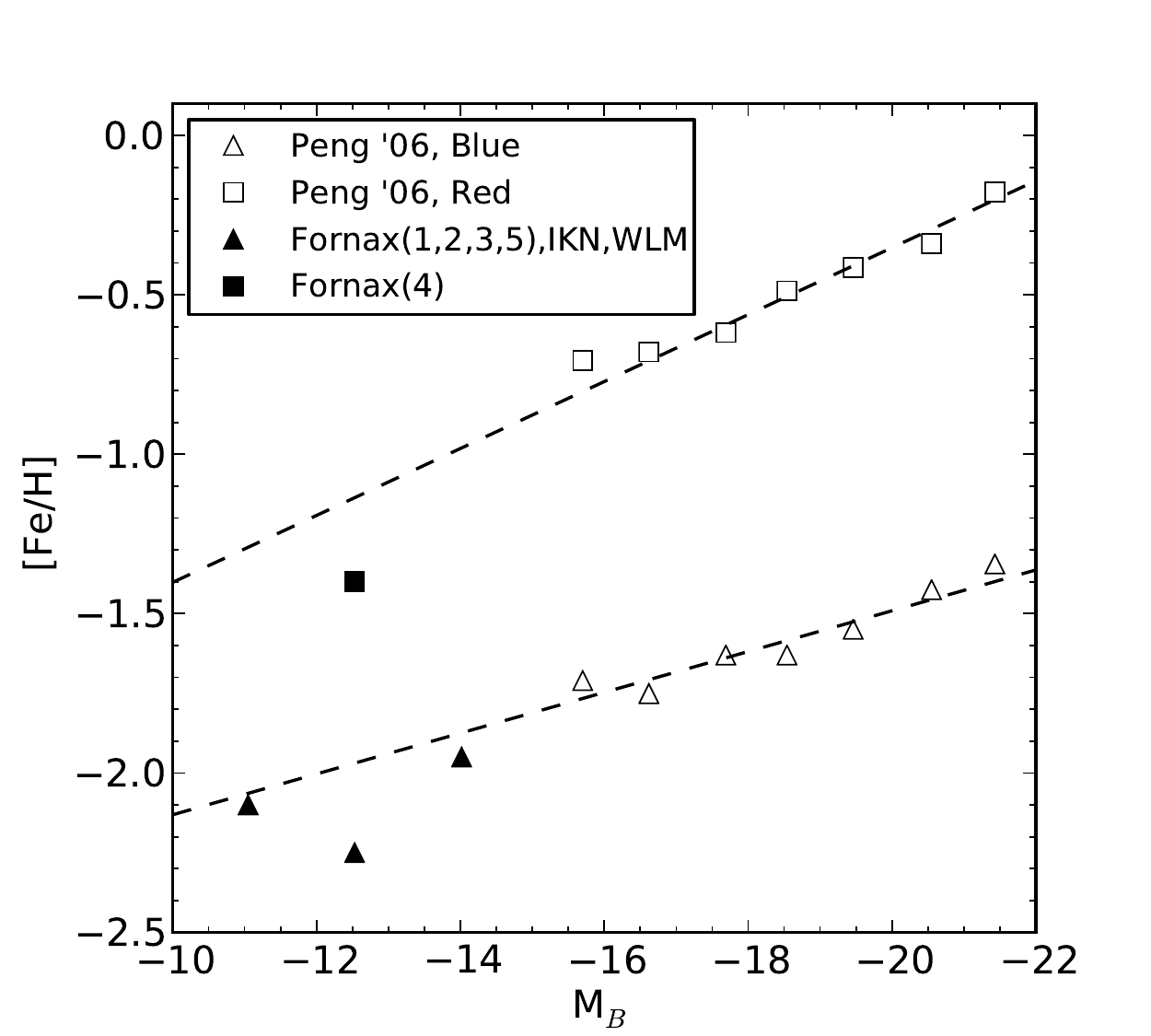}
\caption{\label{fig:mbfeh}Mean metallicities of metal-poor and metal-rich GC sub-populations in galaxies versus host galaxy absolute blue magnitude. Open symbols and dashed lines are data and the best-fitting linear relations from \citet{Peng2006}. Filled symbols are high-dispersion spectroscopic measurements from this paper and \citetalias{Larsen2012a}.}
\end{figure}

In terms of their detailed chemical composition, the GCs in the WLM and Fornax dwarf galaxies appear quite similar to those in the Milky Way with the most significant difference being the lower metallicities of the GCs in the dwarf galaxies.  The Milky Way GC system can be divided into two subpopulations with peak metallicities at $\mathrm{[Fe/H]}\approx-1.5$ and $\approx-0.5$ \citep[e.g.][]{Zinn1985,Roediger2014}. This is quite typical of most large galaxies \citep{Larsen2001,Brodie2006,Peng2006,Usher2012}, although it is still debated to what extent there is always a clean division into two sub-populations  \citep{Yoon2006,Cantiello2007,Chies-Santos2012}. 

A number of studies have found that the metallicities of GC sub-populations are correlated with the masses/luminosities of their parent galaxies, mostly based on broadband photometry of the GC systems around larger galaxies \citep{Larsen2001,Strader2004a,Peng2006}.
Figure~\ref{fig:mbfeh} shows the mean [Fe/H] vs.\ host galaxy $M_B$ relation for blue (metal-poor) and red (metal-rich) GC sub-populations, according to \citet{Peng2006}.  The GC metallicities were derived from Gaussian fits to the $g-z$ colour distributions. The best-fitting straight lines from Peng et al.\ are also included. We have added the dwarf galaxies for which GC metallicities have been measured with high-dispersion spectroscopy, dividing the Fornax GC system into a red ``subpopulation'' consisting of Fornax 4 and a blue subpopulation encompassing the four other clusters. 
The correlation between the metallicity of the metal-poor GCs and the host galaxy luminosity appears to extend to the dwarfs, although it is not clear whether GCs in the dwarf galaxies all fall exactly on the extrapolated relations of Peng et al.
Particularly intriguing is the possibility that Fornax 4 might be the equivalent of the metal-rich GCs in larger galaxies.

Similar to the situation in large galaxies \citep{Forte1981,Forbes1997,Larsen2001,Forbes2001,Harris2002,Harris2007}, the GCs in the WLM, IKN, and Fornax dwarf galaxies are more metal-poor on average than the stellar populations in their parent galaxies \citep[see also][]{Letarte2006}. 
The reason for this is currently unclear but may be related to different formation efficiencies, amounts of disruption, or some combination thereof. It may also be a result of forming the GCs first in a relatively short burst at the beginning of a starburst, so that the stars form over a longer period and reach higher metallicities 
\citep{Harris2002,Brodie2006}. However, this last scenario would have to be tested against spatial distribution constraints when these eventually become available for metal-poor galaxy halos.

\section{Summary and conclusions}

We have presented new VLT/UVES high-dispersion, integrated-light spectroscopy of the globular cluster in the \object{WLM}  galaxy. With these data, we have measured the abundances of several light, $\alpha$, Fe-peak, and $n$-capture elements and compared the results with data for Milky Way GCs and field stars and with literature data for extragalactic GCs in the \object{Fornax dSph} and \object{M31}. We have also determined the metallicity of the brightest GC in the IKN dwarf spheroidal in the M81 group, using a new Keck/ESI spectrum.
Our main findings and conclusions are as follows:

\begin{enumerate}
\item  We measure metallicities of $\mathrm{[Fe/H]}=-1.96\pm0.03$ and $\mathrm{[Fe/H]}\approx-2.1$ for the \object{WLM GC} and \object{IKN-5}. While there may be systematic uncertainties at the level of $\sim0.1$ dex, it is clear that these GCs are both significantly more metal-poor than a typical metal-poor GC in the Milky Way halo. They are also significantly more metal-poor than the average of the field stars in their parent galaxies.
\item By comparison with literature data for the field-star metallicity distributions and star-formation histories, we estimate that the \object{WLM GC} accounts for 17\%--31\% of the metal-poor stars in WLM, while the number of metal-poor stars in the GC \object{IKN-5} may even be comparable to that in the rest of the IKN dwarf galaxy.
This makes these two dwarfs similar to the \object{Fornax dSph} in that they have a very high GC-to-field star ratio at low metallicities.
\item The GCs in the \object{WLM} and Fornax dwarfs generally have enhanced $\alpha$-element abundances at the level of $\approx+0.3$ dex, as traced by the [Ca/Fe] and [Ti/Fe] ratios. The enhanced $\alpha$-element abundances indicate that chemical enrichment was prompt and dominated by Type II SNe nucleosynthesis for the material out of which these clusters formed. The only exception is \object{Fornax 4}, whose lower [$\alpha$/Fe] ratio is similar to that of field stars of similar metallicity in Fornax.
\item The [Mg/Fe] ratios are significantly lower than [Ca/Fe] and [Ti/Fe]. This may be due to anomalous Mg abundances in the cluster stars, although we point out the puzzling observation that the majority of GCs studied in integrated light so far exhibit this phenomenon, while only a small fraction of Milky Way GCs appear to have mean [Mg/Fe] abundances as low as those observed in the  integrated-light studies.
\item The integrated-light [Na/Fe] ratio in the \object{WLM GC} is about 2$\sigma$ higher than in Milky Way field stars and more similar to the typical average [Na/Fe] abundances of Milky Way GCs. This is consistent with the idea that the WLM GC hosts significant numbers of ``second-generation'' stars which formed out of material that had been processed by $p$-capture nucleosynthesis at high temperatures.
\item The Fe-peak elements (Cr, Sc, Mn) and the $n-$capture elements (Ba, Y, La) in the WLM and Fornax GCs generally follow the trends observed in Milky Way field stars and GCs.
\end{enumerate}

Overall, the chemical composition of the \object{WLM GC} and the Fornax GCs is fairly similar to those of globular clusters of corresponding metallicity in the Milky Way, suggesting that these different environments shared relatively similar early chemical enrichment histories. However, the interstellar gas in the dwarf galaxies had reached a lower level of overall chemical enrichment at the time when the majority of the GCs formed, compared to larger galaxies like the Milky Way, and the dwarfs were apparently able to form bound, massive star clusters at extremely high efficiency compared to field stars.

In the context of GC formation scenarios, it is noteworthy that the integrated-light abundance patterns hint at the presence of the same light-element abundance anomalies in the WLM and Fornax clusters (enhanced [Na/Fe] and depleted [Mg/Fe]) that are known from Milky Way GCs. This, combined with the high ratios of GCs vs. field stars in the dwarfs, would appear to favour scenarios for the origin of abundance anomalies within GCs that do not require the clusters to have lost a very large fraction of their initial mass. It also constrains the amount of mass that could have been lost from disrupted star clusters to the field more generally.

\begin{acknowledgements}
 
DF thanks the ARC for their support via DP130100388 and JB acknowledges NSF grant AST-1109878.
We acknowledge  S.\ Penny for help with the ESI observations, J.\ Roediger for discussions about the comparison with Milky Way GCs, and H.\ Zhang for discussions about the mass of WLM. We thank the anonymous referee for a careful and critical reading of the manuscript.  
The Keck Observatory was made possible by the generous financial support of the W.M. Keck Foundation.
This research has made use of NASA's Astrophysics Data System Bibliographic Services.
\end{acknowledgements}

\bibliographystyle{aa}
\bibliography{libmen.bib}

\onecolumn

\end{document}